\documentclass[12pt]{article}
\usepackage{amsfonts}
\usepackage{latexsym}
\usepackage{amsmath}
\usepackage{amssymb}
\usepackage{amssymb}

\newcommand{\newsection}{    
\setcounter{equation}{0}\section}
\def\appendix#1{\addtocounter{section}{1}\setcounter{equation}{0}
\renewcommand{\thesection}{\Alph{section}}
\section*{Appendix \thesection\protect\indent \parbox[t]{11.15cm}{#1}}
\addcontentsline{toc}{section}{Appendix \thesection\ \ \ #1}}

\newcommand{\be}{\begin{eqnarray}}
\newcommand{\ee}{\end{eqnarray}}
\newcommand{\bea}{\begin{eqnarray}}
\newcommand{\eea}{\end{eqnarray}}
\newcommand{\ba}{\begin{array}}
\newcommand{\ea}{\end{array}}

\newcommand{\la}{\label}

\def\a{\alpha}
\def\b{\beta}

\def\e{\epsilon}

\font\mybb=msbm10 at 11pt

\def\bb#1{\hbox{\mybb#1}}

\def\bR {\bb{R}}

\def\bH {\bb{H}}
\def\bC {\bb{C}}

\def\lc{\lrcorner}
\def\ra{\rangle}

\def\ui {{\underline {I}}}
\def\uj {{\underline {J}}}

\def\ua{{\underline {a}}}
\def\ub{{\underline {b}}}

\def\uM{{\underline {M}}}
\def\uN{{\underline {N}}}

\def\uA{{\underline {A}}}
\def\uB{{\underline {B}}}

\def\uo{{\underline {1}}}
\def\u2{{\underline {2}}}

\def\pa{{ {a'}}}
\def\pb{{ {b'}}}
\def\pc{{{c'}}}

\def\pr{{{r'}}}
\def\ps{{{s'}}}
\def\pt{{ {t'}}}

\def\ur{{\underline {r}}}
\def\us{{\underline {s}}}

\begin{document}
\begin{titlepage}
\begin{center}
\vspace{5.0cm}

\vspace{3.0cm} {\Large \bf Spinorial geometry and Killing spinor equations of  6-D supergravity}

\vspace{3.0cm}

{}\vspace{2.0cm}
 {\large
M. Akyol${}^\dagger$ and  G.~Papadopoulos${}^+$\footnote{On study leave from the Department of
Mathematics, King's College London, Strand,
London WC2R 2LS, UK.}
 }

{}

\vspace{0.5cm}
${}^\dagger$
Department of Mathematics,
King's College London\\
Strand,
London WC2R 2LS, UK\\

\vspace{0.5cm}
${}^+$
AEI, Max-Planck Institute f\"ur Gravitationsphysik\\
Am M\"uhlenberg 1,
D-14476 Potsdam\\
Germany\\

\end{center}
{}
\vskip 3.0 cm
\begin{abstract}

We solve the Killing spinor equations of  6-dimensional (1,0)-supergravity coupled to any number of tensor, vector and scalar multiplets
in all cases. The  isotropy groups of Killing spinors are $Sp(1)\cdot Sp(1)\ltimes \bH (1)$, $U(1)\cdot Sp(1)\ltimes \bH (2)$,
$Sp(1)\ltimes \bH  (3,4)$, $Sp(1) (2)$, $U(1) (4)$ and $\{1\} (8)$, where in parenthesis is the number of supersymmetries preserved
in each case. If the isotropy group is non-compact, the spacetime admits a parallel null 1-form with respect to a connection with
torsion the 3-form field strength of the gravitational multiplet. The associated vector field is Killing and the 3-form is determined
in terms of the geometry of spacetime. The $Sp(1)\ltimes \bH$ case admits a descendant solution preserving 3 out of  4 supersymmetries
due to the hyperini Killing spinor equation. If the isotropy group is compact, the spacetime admits a natural frame constructed
from 1-form spinor bi-linears. In the $Sp(1)$ and $U(1)$ cases,  the spacetime admits 3 and 4 parallel 1-forms
with respect to the connection with torsion, respectively. The associated vector fields are Killing and under some additional restrictions
the spacetime is a principal bundle with fibre a Lorentzian Lie group. The conditions imposed by the Killing spinor equations  on all other
fields are also determined.

\end{abstract}
\end{titlepage}

\setcounter{section}{0}
\setcounter{subsection}{0}


\newsection{Introduction}

In the past few years, there has been much work done to systematically solve the Killing spinor equations (KSEs) of supergravity
theories and identify all solutions which preserve a fraction of spacetime supersymmetry.
This programme, apart from its applications
to supersymmetric theories, string theory and black holes, resembles the classification of instantons and monopoles of gauge theories.
The difference is that the spacetime is now curved and there is a connection with special geometric structures on manifolds.

There are several supergravity theories in 6-dimensions. Here we shall be concerned with (1,0) supergravity, 8 real supercharges,
coupled to tensor, vector and scalar multiplets. The theory has been constructed in \cite{sezgin, ferrara, riccioni}. The KSEs of 6-dimensional
supergravities have been solved before in various special cases. In particular, the KSEs of minimal (1,0) supergravity
have been solved in \cite{dario}, and the maximally supersymmetric
backgrounds have been classified in \cite{dario, jose}.
The KSEs of (1,0) supergravity coupled to a tensor and some vector multiplets have been solved for backgrounds preserving
one supersymmetry in \cite{han}. The KSEs of (1,0) supergravity coupled a tensor, some vector and gauge multiplets have been
solved for backgrounds preserving one supersymmetry in  \cite{jong}, see also \cite{gueven}. Most of the computations
carried out so far have been based on the method of spinor bi-linears \cite{hull} first applied to
5-dimensional supergravity. The only exception is the work of \cite{jose} where the integrability conditions
of the KSEs were used as in \cite{gpjose}.

In this paper, we shall solve the KSE of (1,0) supergravity coupled to any number of tensor, vector and scalar multiplets
for backgrounds preserving any number of supersymmetries. For this, we shall use the spinorial geometry method of \cite{spingeom}
and the apparent analogy that exists between the KSEs of (1,0) supergravity and those of heterotic supergravity. The latter have
been solved in all generality  \cite{het1, het2, het3}. We find that the solutions are characterized uniquely, apart from one case,
by the isotropy group of the Killing spinors in $Spin(5,1)\cdot Sp(1)$. This is the holonomy of the supercovariant connection
of a generic background. In particular, the isotropy groups of the spinors are
\bea
&&Sp(1)\cdot Sp(1)\ltimes\bH (1)~,~~U(1)\cdot Sp(1)\ltimes \bH(2)~,~~Sp(1)\ltimes\bH(3,4)~;
\cr
&&Sp(1)(2)~,~~U(1)(4)~,~~\{1\}(8)~,
\eea
where in parenthesis is the number of Killing spinors. Observe that in the
 $Sp(1)\ltimes\bH$ case there is the possibility of a background to admit either 3 or 4 Killing spinors. To explain this,
 we note that in general  only some of the solutions of the gravitino KSE to be also solutions of the
 other KSEs. Backgrounds for which the gravitino admits more solutions than the other KSEs are called descendants, see \cite{het2}.
 In the (1,0) supergravity, all backgrounds for which the gravitino KSE admits 4 or more solutions have descendants. However,
 after an analysis, we have shown that most of the descendants are not independent. This means that most of the descendant
 solutions are special cases of others  for which all solutions of the gravitino KSE
 are also solutions of the other KSEs. The only case that this does not happen is that for the descendant $Sp(1)\ltimes\bH$ backgrounds
 which preserve 3 supersymmetries. As we shall see, the conditions that arise from the hyperini KSE for 3 and 4 supersymmetries
 are different and so the $N=3$ case gives rise to an independent descendant.
 The results on isotropy groups and the analysis for the descendants have been summarized
in tables 1 and 2.

The geometry of the solutions depends on the isotropy group of the Killing spinors. There are two classes of solutions
depending on whether the isotropy group is compact or non-compact. In the non-compact case and for backgrounds preserving one supersymmetry,
the spacetime admits
a parallel 1-form with respect to a metric connection, $\hat\nabla$, with skew-symmetric torsion, $H$, given by the 3-form
field strength of the gravitational multiplet. As a result the spacetime admits a null Killing vector field. The 3-form
field of the gravitational multiplet is completely determined in terms of the geometry of spacetime. In turn, the geometry
of spacetime is characterized by the above mentioned parallel 1-form and a triplet of null 3-forms\footnote{These 3-forms are twisted
with respect to an  $Sp(1)$. So they should be thought of as a vector bundle valued
3-forms.} which are constructed
as Killing spinor bi-linears. The triplet of 3-forms in the directions transverse to the light-cone can be identified
with the Hermitian self-dual forms in 4-dimensions. The 3-forms are also covariantly constant but this time  with respect to
 a connection, ${\cal D}$,
which apart from the skew-symmetric torsion part mentioned above,  also  includes an $Sp(1)$ connection which rotates
the 3-forms.  Such condition  is similar to that of  Quaternionic K\"ahler
 with torsion geometry  \cite{qkt}. The only difference is that the $Sp(1)$ connection may depend
 on the scalars of the hypermultiplet. In the $N=2$ case, the spacetime admits the same form bi-linears, and
 so a null Killing vector field. The main difference is that one of the 3-form bi-linears is now parallel with respect
 to $\hat\nabla$. Though for the other two the covariant constancy conditions involves an additional
  $U(1)$ connection. Similarly in the $N=4$ case, the spacetime admits the same form bi-linears. However
 all the 3-form bi-linears are now parallel with respect to $\hat\nabla$. The geometry
 of solutions with 3 supersymmetries is the same as that of backgrounds which preserve 4 supersymmetries. The difference is in the conditions
 that arise from the hyperini KSE.

 In the compact case and for backgrounds preserving 2 supersymmetries, the spacetime admits 3 parallel 1-forms with respect
 to $\hat\nabla$. Therefore, the spacetime admits 3 isometries and $H$ is determined in terms of these 1-forms and their first derivatives.
 The spacetime also admits 3 additional (vector bundle valued) 1-form bi-linears which now are parallel with respect to
${\cal D}$ connection. Therefore the co-tangent space of spacetime decomposes into a trivial rank 3 bundle spanned
by the $\hat\nabla$-parallel 1-forms  and the rest. Under some additional conditions, which are not implied by the KSEs, the
spacetime can be thought as a principal bundle but in such a case it becomes a product $G\times \Sigma$, where $G$ is locally
$\bR^{3,1}$ or $SL(2,\bR)$ and $B$ is a 3-dimensional Riemannian manifold. The curvature of $B$ is identified
with that of an  $Sp(1)$ connection which may be induced from
the Quaternionic-K\"ahler manifold of scalar multiplets. Next for backgrounds which preserve 4 supersymmetries,
the spacetime admits 4 $\hat\nabla$-parallel 1-form bi-linears. It also admits 2 (vector bundle valued) 1-form bi-linears which now are
parallel with respect to
${\cal D}$ connection. Therefore the spacetime admits at least 4 isometries. The co-tangent spaces decomposes  into a trivial rank 4 bundle spanned
by the $\hat\nabla$-parallel 1-forms  and the rest. As in the previous case, under some additional conditions which are not implied by the KSEs, the
spacetime can be thought as a principal bundle. The fibre group has Lie algebra $\bR^{3,1}$ or $\mathfrak{sl}(2,\bR)\oplus \mathfrak{u}(1)$
or $\mathfrak{cw}_6$. However unlike the previous case, if the fibre group is not abelian, the spacetime is not a product. The curvature
of the base space $B$ is identified with that of a $U(1)$ connection which may be induced from
the Quaternionic-K\"ahler manifold. In both compact and non-compact cases, the conditions imposed on the other fields from the KSEs
 have all been solved. In addition the fields have been expressed in terms of the geometry and their independent components.

This paper has been organized as follows. In section 2, we review the KSEs of 6-dimensional supergravity and explain
their relation to those of heterotic supergravity. In section 3, we describe the solutions of the gravitino KSE and
investigate the existence of descendants. In section 4, we present the geometry of backgrounds preserving 1 supersymmetry.
In sections 5 and 6, we describe the geometry of backgrounds preserving 2 supersymmetries. Similarly in sections
7 and 8, we investigate the geometry of backgrounds preserving 4 supersymmetries as well as that of the $N=3$ descendant.
In section 9,  we describe the backgrounds which preserve all 8 supersymmetries, and in section 10 we give our conclusions.

\newsection{$(1,0)$ supergravity}

\subsection{Fields and KSEs}

There are four types of (1,0)-supersymmetry multiplets in 6 dimensions, the graviton, tensor, vector and scalar
mulptiplets. The bosonic fields of these multiplets are as follows: the graviton multiplet apart from the graviton has a
2-form gauge potential; the tensor multiplet has a 2-form gauge potential and a real scalar; the vector multiplet
has a vector and the scalar multiplet has 4 (real) scalars. The theory we shall consider is (1,0)-supergravity coupled
to $n_T$ tensor, $n_V$ vector and $n_H$ scalar multiplets. The bosonic fields of the scalar multiplet, which is also referred as hypermultiplet,
take values in a Quaternionic K\"ahler manifold which has real dimension $4n_H$.

Before we proceed to describe the KSEs, it is important to note that the fermions that appear in (1,0) supergravity
satisfy a symplectic Majorana condition. This condition utilizes the invariant $Sp(1)$ and $Sp(n_H)$ forms
to impose a reality condition of the spinors. Suppose that the Dirac or Weyl spinors $\lambda$ and $\chi$ transform under the
fundamental representations of $Sp(1)$ and $Sp(n_H)$, respectively. The symplectic Majorana condition is given by
\bea
\lambda^\uA= \epsilon^{\uA\uB} C \bar\lambda^T_\uB~,~~~\chi^\ua= \epsilon^{\ua\ub} C \bar \chi^T_\ub~,
\eea
where $C$ is the charge conjugation matrix and $\epsilon^{\uA\uB}$ and $\epsilon^{\ua\ub}$ are the symplectic invariant
forms of $Sp(1)$ and $Sp(n_H)$, respectively, and $\uA, \uB=1,2$ and $\ua,\ub=1,\dots, 2n_H$.

We write the  supersymmetry transformations of  the fermions evaluated at the bosonic fields as
\bea
\delta\Psi^\uA_\mu&=&\nabla_\mu\epsilon^\uA-{1\over8}  H_{\mu\nu\rho} \gamma^{\nu\rho}\epsilon^\uA
+{\cal C}_\mu{}^\uA{}_\uB\, \epsilon^\uB~,
\cr
\delta \chi^{\uM\uA}&= &{i\over2} T^\uM_\mu \gamma^\mu \epsilon^\uA
-{i\over24} H^\uM_{\mu\nu\rho} \gamma^{\mu\nu\rho} \epsilon^\uA~,
\cr
\delta \psi^\ua&=& i\gamma^\mu \epsilon_\uA  V^{\ua\uA}_\mu ~,
\cr
\delta \lambda^{\pa\uA}&=& -{1\over 2\sqrt 2} F^{\pa}_{\mu\nu} \gamma^{\mu\nu} \epsilon^\uA-{1\over\sqrt{2}} (\mu^{\pa})^\uA{}_\uB  \e^\uB~,
\la{6kse}
\eea
where $\Psi$ is the gravitino, $\chi$ is the tensorini, $\psi$ is the hyperini and $\lambda$ is the gaugini,
$\epsilon$ is the superymmetry parameter and $\pa=1,\dots, n_V$. The remaining coefficients that appear in the supersymmetry
transformations depend on the fundamental fields of the theory. In turn, their explicit expressions depend on the formulation of the theory.
The above structure of the superymmetry transformations that we have stated includes all known formulations.
 Most of the analysis  on the solutions of the KSEs that follows is independent on the precise expression
of  supersymmetry transformations in terms of the fields. Because of this, we shall  give the
conditions that arise from the KSEs in generality. We shall also state explicitly where we use  expression of the
KSEs  in terms of the fields. In what follows, we shall always assume that
$\nabla$ is the spin connection of the spacetime and ${\cal C}$ is a $Sp(1)$ connection.

To give an example of how the supersymmetry transformations,(\ref{6kse}),  depend on the fundamental fields of the theory, we shall mostly
use the formulation\footnote{We use a different normalization for some of the fields from that in \cite{riccioni}. Our normalization
is similar to that of heterotic supergravity.} proposed in \cite{riccioni}. In this formulation, the organization of the fields is as follows.
The theory has $n_T+1$ 2-form gauge potentials
$B^\ur$, $r=0,1,\dots, n_T$.
One of the 2-form potentials is associated  with the gravitational multiplet and the remaining  $n_T$ with the tensor multiplets.
Let us denote the corresponding 3-form field strengths with $G^\ur$. The precise relation between $B^\ur$ and
$G^\ur$ will be given later as well as the duality conditions on $G^\ur$.
To continue, the scalar fields of the tensor multiplets parameterize the coset space $SO(1, n_T)/SO(n_T)$.
 A convenient way to describe this coset space is to choose a local section  $S$  as
\begin{equation}
S= \begin{pmatrix}
v_\ur\\ x_\ur^\uM
 \end{pmatrix}~,~~~\uM=1,\dots n_T
\end{equation}
Since $S\in SO(1, n_T)$, one has $\tilde S\eta S= \eta$ where $\eta$ is the Lorentz metric in $(1, n_T)$-dimensions. In
particular
\bea
v_\ur v^\ur=1~,~~~v_\ur v_\us- \sum_\uM x_\ur^\uM x_\us^\uM=\eta_{\ur\us}~,~~~v^\ur x_\ur^\uM=0~.
\eea
The canonical $SO(n_T)$ connection of the coset is $\sum_\ur x_\ur^\uM d x_\ur^\uN$.

The scalars of the hypermultiplet parameterize a Quaternionic K\"ahler manifold which has holonomy
$Sp(n_H)\cdot Sp(1)$. Such a manifold admits a frame $E$ such that the metric can be written as
\bea
g_{\ui\uj}= E_\ui^{\ua \uA} E_\uj^{\ub \uB} \epsilon_{\ua\ub} \epsilon_{\uA\uB}~,
\eea
where $\epsilon_{\ua\ub}$ and $\epsilon_{\uA\uB}$ are the invariant $Sp(n_H)$ and $Sp(1)$ 2-forms, respectively. The associated
spin connection has holonomy $Sp(n_H)\cdot Sp(1)$ and so decomposes as $({\cal A}_\ui^\ua{}_\ub, {\cal A}_\ui^\uA{}_\uB)$.

In \cite{riccioni} to include vector multiplets with (non-abelian) gauge potential $A_\mu^\pa$, one assumes that the
quaternionic K\"ahler manifold\footnote{It is likely that this assumption is not necessary and a more general class
of models can exist. Moreover $\mu$ may be related to moment maps \cite{gal} of Quaternionic K\"ahler geometry.} of the hypermultiplet
is $Sp(1, n_H)/Sp(1)\times Sp(n_H)$ and gauges the maximal
compact isometry subgroup $Sp(1)\times Sp(n_H)$. So the gauge group of the theory is $H=Sp(1)\times Sp(n_H)\times K$, where $K$
is a product of semi-simple groups which does not act on the scalars. Let $\xi_{a'_1}$ and $\xi_{a'_2}$ be the vector fields
generated on $Sp(1, n_H)/Sp(1)\times Sp(n_H)$ by the action of $Sp(1)$ and $Sp(n_H)$, respectively.
 Under these assumptions, one has that
\bea
&&H_{\mu\nu\rho}=v_\ur G^\ur_{\mu\nu\rho}~,~~~H^\uM_{\mu\nu\rho}=x^\uM_\ur G^\ur_{\mu\nu\rho}~,~~~
{\cal C_\mu}{}^\uA{}_\uB=D_\mu\phi^{\ui} {\cal A}_\ui{}^\uA{}_\uB~,
\cr
&&T^\uM_\mu=x_\ur^\uM\partial_\mu v^\ur~,~~~ V^{\ua\uA}_\mu=E^{\ua\uA}_\ui D_\mu \phi^\ui~,~~~
F^{\pa}_{\mu\nu}= \partial_\mu A_\nu^{\pa}-\partial_\nu A_\mu^{\pa}+ f^{\pa}{}_{\pb\pc} A_\mu^{\pb} A_\nu^{\pc}~,
\cr
&&(\mu^{a'_1})^\uA{}_\uB  = -{2\over v_\ur c^{\ur 1}}
{\cal A}_\ui{}^\uA{}_\uB \xi^{\ui a'_1}~,~~~(\mu^{a'_2})^\uA{}_\uB  = -{2\over v_\ur c^{\ur 2}}
{\cal A}_\ui{}^\uA{}_\uB \xi^{\ui a'_2}~,~~
(\mu^{a'_3})^\uA{}_\uB  =0~,
\la{ric}
\eea
where the gauge index $a'_3$ ranges over the gauge subgroup $K$,  $\phi^\ui$ are the scalars of the hypermultiplet,
\bea
\nabla_\mu\epsilon^\uA&=&\partial_\mu\epsilon^\uA+{1\over4} \Omega_{\mu, mn} \gamma^{mn} \epsilon^\uA~,
\cr
D_\mu \phi^\ui&=&\partial_\mu\phi^\ui- A_\mu^{\pa} \xi_{\pa}^\ui~,
\eea
respectively, and $\Omega$ is the frame connection of spacetime. It is understood that $\xi_{a'_3}=0$
as $K$ does not act on the scalars of the hypermultiplet.
Clearly $F^{\pa}$ are the field strengths of the gauge potentials $A^\pa$
and $f$ are the structure constants of the gauge group $H$.

It remains to define the field strengths $G^\ur$. These are given by
\bea
G^\ur_{\mu\nu\rho} =3 \partial_{[\mu} B^\ur_{\nu\rho]}+ c^{\ur 1} CS(A^{Sp(1)})_{\mu\nu\rho}+ c^{\ur 2}
CS(A^{Sp(n_H)})_{\mu\nu\rho}+ c^{\ur K} CS(A^{K})_{\mu\nu\rho}~,
\eea
where $c^\ur$\,'s are constants, one for each copy of the gauge group, and $CS(A)$'s are the Chern-Simons 3-forms.
Observe that the constants $c^{\ur 1}$ and $c^{\ur 2}$
enter in the definition of $\mu$'s in (\ref{ric}).

The duality condition on $G$ is given by
\bea
\zeta_{\ur\us} G^\us_{\mu_1\mu_2\mu_3}={1\over 3!} \epsilon_{\mu_1\mu_2\mu_3}{}^{\nu_1\nu_2\nu_3} G_{\ur \nu_1\nu_2\nu_3}~,
\eea
where
\bea
\zeta_{\ur\us}= v_\ur v_\us+ \sum_\uM x^\uM_\ur x^\uM_\us~.
\eea
Note that the duality conditions for $H$ and $H^\uM$ are opposite. In our conventions, $H$ is anti-self-dual while $H^\uM$
is self-dual.

\subsection{Spinors}

The spinorial geometry technique to solve the Killing spinor equations is applied most effectively
provided we express the spinors in terms of forms. In particular,  we have to find a way to impose the  symplectic
Majorana condition on the spinors. For this we identify  the symplectic Majorana-Weyl    $Spin(5,1)$   spinors  with
$SU(2)$-invariant Majorana-Weyl $Spin(9,1)$ spinors.  Under this identification the symplectic-Majorana condition
on the $Spin(5,1)$ spinors is replaced by the Majorana condition on the $Spin(9,1)$ spinors. To do this explicitly,  recall that the Dirac spinors
of $Spin(9,1)$ are identified with $\Lambda^*(\bC^5)$, and the positive and negative chirality
spinors are the even and odd degree forms, respectively. The gamma matrices of ${\rm Clif}(\bR^{9,1})$ are given by
\bea
\Gamma_0&=&-e_5\wedge +e_5\lc~,~~~\Gamma_5=e_5\wedge +e_5\lc~,
\cr
\Gamma_i&=&e_i\wedge +e_i\lc~,~~~\Gamma_{i+5}=i(e_5\wedge -e_5\lc)~,~~~i=1,2,3,4~,
\eea
where $e_i$, $i=1,\dots,5$, is a Hermitian basis in $\bC^5$.
The gamma matrices of ${\rm Clif}(\bR^{5,1})$ are identified as
\bea
\gamma_\mu=\Gamma_\mu~,~~~\mu=0,1,2~;~~~~ \gamma_\mu=\Gamma_{\mu+2}~,~~~\mu=3,4,5~.
\eea
Therefore the positive chirality Weyl spinors of $Spin(5,1)=SL(2,\bH)$ are
$\Lambda^{\rm ev}(\bC\langle e_1, e_2, e_5\ra)=\bH^2$.
The symplectic Majorana-Weyl condition of $Spin(5,1)$ is  the Majorana-Weyl condition
of $Spin(9,1)$ spinors, ie
\bea
\epsilon^*=\Gamma_{67} \Gamma_{89} \epsilon~,
\eea
where $\epsilon \in \Lambda^{\rm ev} \bC\langle e_1, e_2, e_5\ra\otimes \Lambda^*\bC\langle e_{34}\ra$. In particular a basis for the
symplectic Majorana-Weyl spinors is
\bea
&&1+e_{1234}~,~~~i(1-e_{1234})~,~~~e_{12}- e_{34}~,~~~i(e_{12}+ e_{34})~,~~~
\cr
&&e_{15}+e_{2534}~,~~~i(e_{15}-e_{2534})~,~~~e_{25}-e_{1534}~,~~~i(e_{25}+e_{1534})~.~~~
\label{smw}
\eea
Observe that the above basis selects the diagonal of two copies of the Weyl representation of $Spin(5,1)$, where
the first copy is $\Lambda^{\rm ev}(\bC\langle e_1, e_2, e_5\ra)$ while the second copy includes the auxiliary
direction $e_{34}$. The $SU(2)$ acting on the auxiliary directions $e_3$ and $e_4$ leaves the basis invariant.

\subsection{KSEs revisited}

It remains to rewrite the KSEs of 6-dimensional supergravity in terms of the 10-dimensional
notation we have introduced above. For this, we define $\rho^{\pr}$, $\pr=1,2,3$, such that
\bea
\rho^{1} =  \frac{1}{2}(\Gamma_{38}+\Gamma_{49})~,~~~\rho^{2}= \frac{1}{2}(\Gamma_{89}-\Gamma_{34})~,~~~
\rho^{3}  =  \frac{1}{2}(\Gamma_{39}-\Gamma_{48})~.
\label{spgen}
\eea
Observe that these are the generators of the Lie algebra $Sp(1)$ as it acts on the basis (\ref{smw}). Using this
the KSEs can be rewritten as
\bea
{\cal D}\epsilon\equiv \big(\nabla_{\mu}-{1\over8} H_{\mu\nu\rho} \gamma^{\nu\rho}+ {\cal C}_\mu^{\pr} \rho_{\pr}\big)\epsilon & = & 0,
\cr
\left({i\over2} T^\uM_\mu \gamma^\mu-\frac{i}{24}H_{\mu\nu\rho}^{\uM}\gamma^{\mu\nu\rho}\right)\epsilon & = & 0,
\cr
i\gamma^\mu \epsilon_\uA  V^{\ua\uA}_\mu &=&0~,
\cr
\left({1\over4}F_{\mu\nu}^{\pa}\gamma^{\mu\nu}+{1\over2}\mu_{\pr}^{\pa}\rho^{\pr}\right)\epsilon
& = & 0~.
\la{kkk}
\eea
In the hyperini KSE, it should be understood that
\bea
\epsilon_1=-\epsilon^2~,~~~\epsilon_2=\Gamma_{34} \epsilon^1~,
\eea
where $\epsilon^1$ and $\epsilon^2$ are the components of $\epsilon$ in the two copies of the Weyl representation used
to construct the symplectic-Majorana representation.

\newsection{Parallel  and Killing spinors}

\subsection{Parallel spinors}

The (reduced) holonomy\footnote{We assume that the backgrounds are simply connected or equivalently we consider the universal cover.}
of 6-dimensional  supergravity  supercovariant connection ${\cal D}$, (\ref{kkk}),
 is contained in $Spin(5,1)\cdot Sp(1)$. This is the same as the gauge group of the theory.
 Therefore there are two possibilities. Either
the parallel spinors have a trivial isotropy group in $Spin(5,1)\cdot Sp(1)$  or
the parallel spinors have a non-trivial isotropy group in $Spin(5,1)\cdot Sp(1)$. To investigate the two cases,
consider the integrability of the gravitino Killing spinor equation which gives
\bea
{1\over4} \hat R_{\mu\nu, \rho\sigma} \gamma^{\rho\sigma} \epsilon+
{\cal F}_{\mu\nu}^{\pr} \rho_{\pr}\epsilon=0~,
\la{intcon}
\eea
where
\bea
{\cal F}_{\mu\nu}^{\pr}=\partial_\mu {\cal C}^\pr_\nu-\partial_\nu{\cal C}^\pr_\mu+2 \epsilon^{\pr}{}_{\ps\pt} {\cal C}_\mu^{\ps}
{\cal C}_\nu^{\pt}- H^\lambda{}_{\mu\nu} {\cal C}^\pr_\lambda~,
\eea
and $\hat R$ is the curvature of the connection, $\hat\nabla$, with skew-symmetric torsion $H$ defined as
\bea
\hat\nabla_\mu Y^\nu=\nabla_\mu Y^\nu+{1\over2} H^\nu{}_{\mu\lambda} Y^\lambda~.
\eea

\subsubsection{Trivial isotropy group}

Now if the isotropy group of the parallel spinors is $\{1\}$, a
direct inspection of  (\ref{intcon}) reveals that \bea \hat
R=0~,~~~~{\cal F}=0~. \eea The spacetime is parallelizable with
respect to a connection with skew-symmetric torsion and admits 8
parallel spinors. Moreover,
 the torsion is anti-self-dual. All such spacetimes are group manifolds with anti-self-dual structure constants.

\subsubsection{Non-trivial isotropy group}

Next suppose that the parallel spinors have a non-trivial isotropy group in $Spin(5,1)\cdot Sp(1)$. To find the isotropy
groups, we first remark that $Spin(5,1)=SL(2, \bH)$ and the action of $Spin(5,1)\cdot Sp(1)$
on the symplectic Majorana-Weyl
spinors can be described in terms of quaternions. In particular, the symplectic Majorana-Weyl spinors can be
identified with $\bH^2$ with $Spin(5,1)=SL(2, \bH)$ acting from the left with  quaternionic matrix multiplication while
$Sp(1)$ acts on the right with the conjugate quaternionic multiplication.
Using, this it is easy to see
that there is a single non-trivial orbit of $Spin(5,1)\cdot Sp(1)$ on the symplectic Majorana-Weyl spinors
with isotropy group $Sp(1)\cdot Sp(1)\ltimes \bH$. To continue, we have to determine the action of $Sp(1)\cdot Sp(1)\ltimes \bH$
on $\bH^2$. Decomposing $\bH^2=\bR\oplus {\rm Im} \bH\oplus \bH$, where $\bR$ is chosen to be along the first invariant
spinor, then the action of the isotropy group is
\bea
{\rm Im} \bH\oplus \bH\rightarrow a {\rm Im} \bH\bar a \oplus b \bH \bar a~,
\la{act}
\eea
where $(a,b)\in Sp(1)\cdot Sp(1)$ and $\bar a$ is the quaternionic conjugate of $a\in Sp(1)$.
There are two possibilities. Either the second invariant spinor lies in
${\rm Im} \bH$ or in $\bH$. It cannot lie in both because if there is a non-trivial component in $\bH$, there is
a $\bH$ transformation in $Sp(1)\cdot Sp(1)\ltimes \bH$ such that the component in ${\rm Im} \bH$ can be set to zero.
Now if the second spinor lies in ${\rm Im} \bH$, the isotropy group is $U(1)\cdot Sp(1)\ltimes \bH$. On the other
hand if it lies in $\bH$, the isotropy group is $Sp(1)$. This concludes the analysis for two invariant spinors.

To continue, it is easy to see that if there are 3 invariant spinors, then there always exist an additional one. For 4 invariant spinors,
there are two cases to consider with non-trivial isotropy group. Either all four invariant spinors span
the first copy of $\bH$ in $\bH^2$ and the isotropy group is $Sp(1)\ltimes \bH$, or 2 lie in the first copy
and the other 2  lie in the second copy of $\bH$ in $\bH^2$ and the isotropy group is $U(1)$.
The isotropy group of more than 4 linearly independent spinors is  $\{1\}$. The above results as well as representatives
of the invariant spinors have been summarized in table 1.

\begin{table}[ht]
 \begin{center}
\begin{tabular}{|c|c|c|}
\hline
$N$&${\mathrm{Isotropy ~Groups}}$  & ${\mathrm{Spinors}}$ \\
\hline
\hline
$1$  & $Sp(1)\cdot Sp(1)\ltimes \bH$ & $1+e_{1234}$\\
\hline
$2$  & $(U(1)\cdot
Sp(1))\ltimes\bH$ & $1+e_{1234}~, ~i(1-e_{1234})$\\
\hline
$4$  & $
Sp(1)\ltimes \bH$ & $1+e_{1234}~, ~i(1-e_{1234})~,~e_{12}-e_{34}~,~i(e_{12}+e_{34})$\\
\hline
\hline
$2$  & $
Sp(1))$ & $1+e_{1234}~, ~e_{15}+e_{2345}$\\
\hline
$4$  & $
U(1)$ & $1+e_{1234}~,~i(1-e_{1234})~, ~e_{15}+e_{2345}~,~ i(e_{15}-e_{2345})$\\
\hline
\end{tabular}
\end{center}
\label{ttt}
\caption{\small
The first column gives the number of invariant spinors, the second column the associated isotropy groups
and the third representatives of the invariant spinors. Observe that if 3 spinors are invariant, then there is a fourth one.
Moreover the isotropy group of more than 4 spinors is the identity.}
\end{table}

\subsection{Descendants}

A distinguished class of supersymmetric backgrounds are those for which all parallel spinors given in table 1 are Killing,
ie they solve all KSEs. However,
it is not always the case that all solutions of the gravitino KSE are also solutions of the other three KSEs.
Typically, only some or linear combinations  of the parallel spinors are Killing.  This is similar
to the heterotic case where an extensive analysis was required to identify the ``descendant'' solutions \cite{het2},
ie the solutions that had less Killing  than parallel spinors. However
unlike the heterotic, the analysis required  to identify the descendants backgrounds of 6-dimensional supergravity is  simpler.
As we shall see there are many descendants but in most cases the Killing spinors of descendants are given
in terms of the parallel spinors of table 1. Such descendant backgrounds are special cases of solutions
for which all parallel spinors are Killing.  The objective of the analysis which follows is to find whether
there are backgrounds which have Killing spinors that differ from those given in table 1. If they exist,
such backgrounds will be called independent descendant solutions or simply ``independent''.

In all cases, if a solution has just one Killing spinor, irrespective of the number of parallel spinors,
it is always possible to rotate it so that it is identified with $1+e_{1234}$.   Therefore such descendant backgrounds are included in those
for which $1+e_{1234}$ is both parallel and  Killing spinor and so they are not  independent. Using this, the cases
we have to examine are those  with two or more Killing  and with four or more
parallel spinors.

\subsection{Descendants of four parallel spinors}
\la{d4ps}

\subsubsection{$Sp(1)\ltimes \bH$}

 Suppose that a solution has 4 parallel but only 2 Killing spinors. There are two cases
 to consider depending on the isotropy group of the parallel spinors. If the isotropy group
 of the parallel spinors is $Sp(1)\ltimes \bH$, then the sigma group \cite{het2} is $Spin(1,1)\times Sp(1)\cdot Sp(1)$. The subgroup
 $Sp(1)\cdot Sp(1)=SO(4)$  acts
 with the vector representation on the 4 parallel spinors. In such a case, it is always possible to arrange
 such that the first two Killing spinors are
 \bea
1+e_{1234}~, ~i(1-e_{1234})~.
 \eea
Therefore such solutions are special cases of backgrounds with 2 supersymmetries associated with 2 parallel spinors
 with isotropy group $U(1)\cdot
Sp(1))\ltimes\bH$, and so they are not independent.

Next suppose that a solution has 3 Killing spinors. Again since the subgroup $Sp(1)\cdot Sp(1)$ of the sigma group
acts with the the vector representation, it is always possible to choose the 3 Killing spinors as
\bea
1+e_{1234}~, ~i(1-e_{1234})~,~~~e_{12}-e_{34}~.
\la{dessp1}
\eea
It turns out that if the gravitino, tensorini and gaugini KSE admit (\ref{dessp1}) as a solution, then
they admit also $i(e_{12}+e_{34})$ as a solution. Thus all the parallel spinors of this case solve the three out of four KSEs.
It remains to investigate the hyperini KSE. We shall see that the conditions that arise from the hyperini KSE evaluated
on (\ref{dessp1}) are different from those that one finds when the same KSE is evaluated on all 4 $Sp(1)\ltimes \bH$-invariant
spinors. As a result, the KSEs allow for backgrounds with 3 supersymmetries. However the existence of such backgrounds
depends also on the field equations.

\subsubsection{$U(1)$}

It remains to investigate the case for which the 4 parallel spinors
have isotropy group $U(1)$. The sigma group \cite{het2} in this case is $Spin(3,1)\times U(1)$.  One way to see this is to treat the
directions 2,3 and 4 in the $U(1)$-invariant spinors given in table 1 as auxiliary and suppress them.
Then the spinors can be identified with the Majorana spinors
of $Spin(3,1)$. The $U(1)$ subgroup of the sigma group is generated by spin transformations along the auxiliary directions.
The analysis of the orbits of the sigma group is identical to that of the gauge group of
 4-dimensional supergravity \cite{4d}.
Thus  there are two different cases of descendants with 2 supersymmetries that we must consider. Using in addition the $U(1)$
subgroup of the sigma group, one can arrange such that the
Killing spinors of the two cases are identical to the parallel spinors of table 1 with
isotropy groups $U(1)\cdot Sp(1)\ltimes \bH$
and $Sp(1)$, respectively. Therefore
both cases are special cases of other backgrounds with less parallel spinors, and so they are not independent.

Next consider the case of backgrounds with 3 Killing spinors. The existence
of such backgrounds depends on the details of the Killing spinor equations. To see whether such solution
can exist, one can pick the 3-plane of Killing spinors by using the sigma group to bring  the normal spinor of the 3-plane to a canonical form.
The procedure is explained in detail in \cite{n31, het2}. It turns out that the  normal spinor can be chosen such that the 3 Killing spinors lie
on the 3-plane spanned by
\bea
1+e_{1234}~,~i(1-e_{1234})~, ~e_{15}+e_{2345}~.
\la{desu1}
\eea
It is easy to see that  if (\ref{desu1}) solve the gravitino, tensorini and gaugini KSEs, then $i(e_{15}-e_{2345})$  is also a solution.
As a result all 4 $U(1)$-invariant spinors are solutions to these three KSEs. It remains to examine the hyperini KSE. Unlike the previous case,
the hyperini KSE evaluated on (\ref{desu1}) gives the same conditions as those one obtains
for all 4 $U(1)$-invariant spinors. Thus in this case there are no descendants preserving 3
supersymmetries.

\begin{table}[ht]
 \begin{center}
\begin{tabular}{|c|c|}\hline
   ${\rm hol}({\cal D})$ &$N$
 \\ \hline \hline
  $Sp(1)\cdot Sp(1)\ltimes\bH$& 1 \\
\hline
$U(1)\cdot Sp(1)\ltimes\bH$&$*$, 2
\\ \hline
$Sp(1)\ltimes\bH$&$*$, $*$, $3$, 4
\\ \hline \hline
$Sp(1)$&$*$, 2
\\ \hline
$U(1)$&$*$, $*$, $-$, 4
\\ \hline
$\{1\}$& $*$,$*$,$*$, $*$,$-$, $-$, $-$, 8
\\ \hline
\end{tabular}
\end{center}
\caption{\small
In the  columns are the holonomy  groups that arise from the solution of the gravitino KSE
and the number $N$  of supersymmetries, respectively. $*$ entries denote the cases that occur
but are special cases of others with the same number of supersymmetries but with less parallel spinors. The $-$  entries denote cases which
do not occur. The Killing spinors for $N=1,2,4$  are the same as those given in table 1 while for $N=3$ in  (\ref{dessp1}).}
\end{table}

\subsection{Descendants of eight parallel spinors}
\la{d8s}

It remains to examine the descendants of backgrounds with 8 parallel spinors.  For this it is convenient to solve the KSEs
in the order
\bea
\mathrm {gravitino}\rightarrow \mathrm {gaugini}\rightarrow \mathrm {tensorini}\rightarrow \mathrm {hyperini}~.
\eea
We have already stated that the gravitino KSE admits 8 parallel spinors. It remains to investigate the remaining three KSEs.

\subsubsection{Gaugini}

The solutions of the gaugini KSE are spinors
which are invariant under some subgroup of $Spin(5,1)\cdot Sp(1)$. This is because the gauge field and moment maps can be viewed
as maps from $\mathfrak{spin}(5,1)\oplus \mathfrak{sp}(1)$ to the Lie algebra of the gauge group, where $\mathfrak{spin}(5,1)=\Lambda^2(\bR^{5,1})$.
But all such spinors and their isotropy groups have been tabulated in table 1.
Thus the gaugini KSE can preserve 1, 2(2), 4(2) and 8 out of the total of 8 parallel spinors, where the number in the
parenthesis states the multiplicity of each case.

Having established that the gaugino KSE has solutions given by the spinors of table 1,
it remains to investigate the remaining two KSEs. If the gaugini KSE has up to 4 solutions, the investigation of the descendants
for the tensorini and hyperini KSEs is the same as that presented in section \ref{d4ps}.  In particular,  there is  one
descendant with 3 supersymmetries
which arises in the case of 4 Killing spinors with isotropy group $Sp(1)\ltimes \bH$. The three
Killing spinors are given in (\ref{dessp1}). So this case can be thought as a special case of backgrounds
with 4 parallel spinors and Killing spinors given in (\ref{dessp1}). Since we have dealt with all descendants of the
gaugini KSE from now on we shall take that the gaugini KSE preserves all 8 parallel spinors.

\subsubsection{Tensorini}

Let us assume that the gravitino and gaugini KSEs  admit 8 Killing spinors. Observe that the tensorini KSE commutes
with all 3 $\rho$ operations given in (\ref{spgen}). Because of this it preserves either 4 or 8 supersymmetries.
Moreover, whenever it preserves 4 supersymmetries, the Killing spinors can be given in terms of the $Sp(1)\ltimes \bH$-invariant
spinors of table 1. Using this, one can solve the hyperini KSE to find that the backgrounds preserve 1,2,3 and 4 supersymmetries.
All of them are special cases
of solutions which we have already investigated. In particular, if the solutions preserves one supersymmetry, then it is a special case
of backgrounds with one parallel spinor which is also Killing. In the N=2 case, the backgrounds are special cases
of solutions with two parallel spinors which are also Killing and have isotropy group $Sp(1)\cdot U(1)\ltimes \bH$. For  $N=3$,
the backgrounds are special cases of those with $Sp(1)\ltimes \bH$-invariant parallel spinors and 3 Killing
spinors given in (\ref{dessp1}). The $N=4$ case is included in that for which the 4  $Sp(1)\ltimes \bH$-invariant parallel spinors are also Killing.
This concludes the analysis of the descendants in this case, so from now one we shall assume that the tensorini KSE admits
8 Killing spinors.

\subsubsection{Hypernini}

Let us assume that the gravitino, gaugini and tensorini KSEs  admit 8 Killing spinors. To investigate solutions of the hyperini KSE,
we have to identify the orbits of the sigma group, which in this case is $Spin(5,1)\cdot Sp(1)$, on the space of spinors.
We have already dealt with the descendants preserving one supersymmetry. The Killing spinor can be identified with $1+e_{1234}$.
To investigate the case with 2 supersymmetries, we first recall
that the sigma group $Spin(5,1)\cdot Sp(1)$ has one orbit in the space
of symplectic-Majorana spinors with isotropy group $Sp(1)\cdot Sp(1)\ltimes \bH$. The representative can be chosen as
$1+e_{1234}$. The action of the isotropy group on the space of spinors
is given in (\ref{act}). This isotropy group has two non-trivial orbits on the space of spinors and the representatives
can be chosen as
 either $i(1-e_{1234})$ or $e_{15}+e_{2345}$.  It is clear from this that solutions with  Killing spinors $1+e_{1234}$ and
 $i(1-e_{1234})$ or $1+e_{1234}$ and $e_{15}+e_{2345}$
 are not independent descendants. So there no independent descendants with two supersymmetries.

 Next let us consider the case with 3 supersymmetries. There are two cases to investigate. First suppose that the isotropy
 group of the first two spinors is $Sp(1)\cdot U(1)\ltimes \bH$. This group has two different orbits on the rest of the spinors
 with representatives
 $e_{12}-e_{34}$ and $e_{15}+e_{2345}$, respectively. These two cases are not new as the Killing spinors are identical
 to those found in (\ref{dessp1}) and (\ref{desu1}), respectively. In addition one can show that if the
 hyperini KSE admits (\ref{desu1}) as Killing spinors, then it preserves 4 supersymmetries with Killing spinors as the $U(1)$-invariant
 spinors of table 1.

Next suppose that the isotropy group of the first two Killing spinors is $Sp(1)$. It can be easily seen from
(\ref{act}) that $Sp(1)$ acts with two copies of the 3-dimensional representation on the remaining 6 spinors.
As a result it can be arranged such that the third spinor can be chosen in such a way that the three Killing spinors are
\bea
1+e_{1234}~,~~~e_{15}+e_{2345}~,~~~c_1 i(1-e_{1234})+ic_2 (e_{15}-e_{2345})+c_3 (e_{25}-e_{1345})~,
\la{des8u1}
\eea
where $c$'s are constants. If $c_1=0$, then the third spinor can be simplified further by choosing $c_3=0$. As we shall see,
there are no new descendants.  The hyperini KSE evaluated on the above spinors implies that either it preserves four
supersymmetries with Killing  spinors   as the $U(1)$-invariant spinors of table 1 or it preserves
all 8 supersymmetries. This depends on the coefficients $c$.

It remains to investigate descendants with 4 supersymmetries. First suppose that the first 3 Killing spinors
are chosen as in (\ref{dessp1}). The isotropy group in this case is $Sp(1)\ltimes \bH$. This has two orbits on the remaining
spinors. The representatives can be chosen such that the four Killing spinors are  given by either the 4 $Sp(1)\ltimes \bH$-invariant
spinors of table 1 or
\bea
1+e_{1234}~,~~~i(1-e_{1234})~,~~~e_{12}-e_{34}~,~~~e_{15}+e_{2345}~.
\eea
This can be a new descendant.  However it turns out that if the hyperini KSE preserves the above 4 spinors, then
it preserves all 8 supersymmetries.

Next suppose that the first 3 Killing spinors are given in (\ref{desu1}). The isotropy group of these spinors is $U(1)$. Thus the fourth
spinor can be chosen as
\bea
c_1 (e_{12}-e_{34})+c_2 i (e_{15}-e_{2345})+c_3 (e_{25}-e_{1345})+ c_4 i (e_{25}+e_{1345})~.
\eea
It turns out depending on the choice of the coefficients $c$ that the hyperini KSE preserves either 4 supersymmetries
with Killing spinors given by the $U(1)$-invariant spinors of table 1 or all 8 supersymmetries. So there are no new descendants.
A similar conclusion holds for the case for which the third Killing spinor is chosen as in (\ref{des8u1}).

To conclude, if the isotropy group of parallel spinors is $\{1\}$, there are descendant backgrounds which preserve 1, 2, 3 and 4
supersymmetries. However they are not independent. All of them are special cases of backgrounds that admit less parallel spinors. The results for all
descendants have been tabulated in table 2.


\newsection{N=1}

The lexicographic structure
of 6-dimensional supergravity KSEs is similar to that of heterotic supergravity. As a result, the
results of \cite{het1, het2} can be  adapted to 6-dimensions.
Because of this, we shall not explain the calculations in detail.
The only difference is in the hyperini KSE which is examined separately.

\subsection{Gravitino}

As the gauge group of the theory is the same as the holonomy of supercovariant connection of generic backgrounds,
the Killing spinor of $N=1$ backgrounds  can be chosen as $\epsilon=1+e_{1234}$, see  \cite{het1, het2} for an explanation. The gravitino KSE
requires that this spinor is parallel. As a result the holonomy of ${\cal D}$ reduces to a subgroup of the isotropy
group  $Sp(1)\cdot Sp(1)\ltimes \bH$ of the parallel spinor, ie
\bea
\mathrm{hol}({\cal D})\subseteq Sp(1)\cdot Sp(1)\ltimes \bH~.
\eea
This is the full content of the gravitino KSE. The restrictions that this imposes on the geometry will be examined later.

\subsection{Gaugini}

A direct application of the spinorial geometry technique \cite{spingeom} reveals that the conditions that arise from the
gaugini KSE are
\bea
F^{\pa}_{+i}=F^{\pa}_{+-}=0~,~~~F^{\pa}_\a{}^\a+i\mu^1=0~,~~~2F^{\pa}_{12}+\mu^2-i\mu^3=0~.
\la{1g1}
\eea
It is clear that the gauge field strength vanishes along one of the light-cone directions.

\subsection{Tensorini}

A direct computation of the tensorini KSE on the spinor $1+e_{1234}$, or a comparison with the
solution of the dilatino KSE for heterotic backgrounds
preserving one supersymmetry,  reveals that
\bea
T^\uM_+=0~,~~~H^\uM_{+\alpha}{}^\alpha=H^\uM_{+\a\b}=0~,
\cr
T^\uM_{\bar\alpha}-{1\over2} H^\uM_{-+\bar\alpha}-{1\over2} H^\uM_{\bar\alpha\beta}{}^\beta=0~.
\la{1tens1}
\eea
Note that the tensorini KSE commutes with the Clifford algebra operations $\rho^{\pr}$ in (\ref{spgen}). As a result, if the tensorini KSE admits a solution $\epsilon$,
then $\rho^{\pr}\epsilon$ also solve the KSE.  As a result, the four spinors
\bea
1+e_{1234}~,~~~\rho^{\pr}(1+e_{1234})~,~~~\pr=1,2,3,
\eea
are  solutions to the tensorini KSE.

\subsection{Hyperini}
To understand the hyperini KSE, one has to identify the $\epsilon_\uA$ components of the Killing spinor
in the context of spinorial geometry. In our notation $\epsilon^1=1$ and $\epsilon^2=e_{1234}$ and since
$\epsilon_1=-\epsilon^2$ and $\epsilon_2=\Gamma_{34}\epsilon^1$, one has $\epsilon_1=-e_{1234}$ and $\epsilon_2=e_{34}$. Substituting
these into the KSE, one finds the conditions
\bea
V_+^{\ua\uA}=0~,~~~-V_1^{\ua\uo}+ V_{\bar2}^{\ua\u2}=0~,~~~ V_2^{\ua\uo}+ V_{\bar1}^{\ua\u2}=0~.
\la{1hyp1}
\eea
Expressing the coefficients of the KSEs in terms of the fundamental fields as in (\ref{ric}),
it is clear that
\bea
D_+\phi^\ui=0~.
\la{dpp}
\eea

\subsection{Geometry}

\subsubsection{ Form spinor bi-linears}

To investigate further the geometry of spacetime, one has to compute the form spinor bi-linears. The form spinor bi-linears
of two  spinors are given by
\bea
\tau={1\over k!} B(\epsilon_1, \gamma_{\mu_1\dots \mu_k} \epsilon_2)\,\, e^{\mu_1}\wedge\dots \wedge e^{\mu_k}~,
\eea
where $B$ is the Majorana inner product as for the heterotic supergravity. Assuming that $\epsilon_1$ and $\epsilon_2$ satisfy the
gravitino KSE, it is easy to see that
\bea
\hat\nabla_\nu \tau=0~.
\eea
The form $\tau$ is covariantly constant with respect to $\hat\nabla$ and
the  $Sp(1)$ connection ${\cal C}^{\pr}$ does not contribute in the parallel transport equation.

On the other hand, one may also consider the $\mathfrak{sp}(1)$-valued form bi-linears
\bea
\tau^{\pr}={1\over k!} B(\epsilon_1, \gamma_{\mu_1\dots \mu_k}\rho^{\pr} \epsilon_2)\,\,
 e^{\mu_1}\wedge\dots \wedge e^{\mu_k}~.
 \eea
 Assuming again that $\epsilon_1$ and $\epsilon_2$ satisfy the gravitino KSE, one finds that
 \bea
 \hat\nabla_\nu \tau^{\pr}+2\, {\cal C}^{\ps}_\nu \epsilon^{\pr}{}_{\ps\pt} \tau^{\pt}=0~.
 \eea
Observe that the $\mathfrak{sp}(1)$-valued form bi-linears are twisted with respect to the  $Sp(1)$ connection ${\cal C}^{\pr}$.
So $\nabla_\nu \tau^{\pr}$ are not forms but rather vector bundle valued forms. However for simplicity in what follows, we shall refer to both
$\tau$ and $\tau^\pr$ as forms.

\subsubsection{Spacetime geometry of N=1 backgrounds}

The algebraic independent bi-linears of backgrounds preserving one supersymmetry are
\bea
e^-~,~~~~e^-\wedge \omega_I~,~~~e^-\wedge \omega_J~,~~~e^-\wedge \omega_K~,
\eea
where $e^-$ is a null one-form and
\bea
\omega_I=-i\delta_{\a\bar\b} e^\a\wedge e^{\bar\b}~,~~~\omega_J=-e^1\wedge e^2-e^{\bar 1}\wedge e^{\bar 2}~,~~~
\omega_K=i(e^1\wedge e^2-e^{\bar 1}\wedge e^{\bar 2})~.
\eea
Clearly $\omega_I, \omega_J$ and $\omega_K$ are Hermitian forms in the directions
transverse to the light-cone. In what follows, we also set $\omega^1=\omega_I$, $\omega^2=\omega_J$ and $\omega^3=\omega_K$.

The conditions that the gravitino KSE imposes on the spacetime geometry can be rewritten as
\bea
\hat\nabla_\mu e^-=0~,~~~\hat\nabla_\mu (e^-\wedge \omega^{\pr})+ 2\,
{\cal C}_\mu^{\ps} \epsilon^{\pr}{}_{\ps \pt} (e^-\wedge\omega^{\pt})=0~.
\la{par1}
\eea
The second equation can be thought as the Lorentzian analogue of the Quaternionic K\"ahler with torsion condition
of \cite{qkt}.
The integrability conditions to these parallel transport equations are
\bea
\hat R_{\mu_1\mu_2, + \nu}=0~,~~~-\hat R_{\mu_1\mu_2,}{}^k{}_i \omega^{\pr}{}_{kj}+(j,i)+2{\cal F}^{\ps}_{\mu_1\mu_2}
\epsilon^{\pr}{}_{\ps\pt} \omega^{\pt}_{ij}=0~.
\la{1int1}
\eea
In addition to this, the torsion $H$ has to be anti-self-dual in 6 dimensions. The conditions for this can be written
as
\bea
H_{+\a\b}=H_{+\a}{}^{\a}=0~,~~~H_{-+\bar\a}+H_{\bar\a\b}{}^\b=0~,~~~H_{-1\bar 1}-H_{-2\bar2}=0~,~~~H_{-1\bar2}=0~,
\la{self1}
\eea
where $\epsilon_{-+1\bar1 2\bar2}=\epsilon_{013245}=-1$.
Notice that from the 4-dimensional perspective, $H_{+ij}$ is an anti-self-dual while $H_{-ij}$ is a self-dual 2-form,
respectively.

To specify the spacetime geometry, one has to solve (\ref{par1}) subject to (\ref{self1}). For this adapt
a frame basis on the spacetime such that one of the light-cone frames is the parallel 1-form $e^-$, ie
the metric is written as
\bea
ds^2=2 e^- e^++\delta_{ij} e^i e^j~.
\la{1metr}
\eea
The first condition in (\ref{par1}) implies that the dual vector field $X$ to $e^-$ is {\it Killing} and
\bea
de^-=i_X H~.
\la{dexh}
\eea
From this, it is easy to see that the torsion 3-form can be written as
\bea
H=e^+\wedge de^-+{1\over2} H_{-ij} e^-\wedge e^i\wedge e^j+ \tilde H~,~~~\tilde H={1\over3!} \tilde H_{ijk} e^i\wedge e^j\wedge e^k~.
\la{1tor}
\eea
Anti-self-duality of $H$ relates the $\tilde H$ component to $de^-$. In particular, one has that
\bea
\tilde H=-{1\over3!}(de^-)_{-\ell}\,\,\epsilon^\ell{}_{ijk} \,\, e^i\wedge e^j\wedge e^k~.
\la{thde}
\eea

This solves the first condition in (\ref{par1}). To solve the remaining three conditions, consider first the
parallel transport equation in (\ref{par1}) along the light-cone directions. Since $H_{+ij}$ is anti-self-dual, one has that
\bea
{\cal D}_+\omega^{\pr}=\nabla_+\omega^{\pr}+ 2\,
{\cal C}_+^{\ps} \epsilon^{\pr}{}_{\ps \pt} \omega^{\pt}=0~.
\la{1geom1}
\eea
This is a condition can be used to express ${\cal C}_+$ in terms of the geometry of spacetime.
Next
\bea
{\cal D}_-\omega^{\pr}_{ij}=\nabla_-\omega^{\pr}_{ij}-H_-{}^k{}_{[i} \omega^{\pr}_{j]k}+
2\, {\cal C}_-^{\ps} \epsilon^{\pr}{}_{\ps\pt} \omega^{\pt}_{ij}=0~.
\eea
Since $H_{-ij}$ is self-dual, this implies that it can be written as
\bea
H_{-ij}=w_{\pr} \omega^{\pr}_{ij}~,
\eea
for some functions $w_{\pr}$.
Thus
\bea
\nabla_-\omega^{\pr}_{ij}+
 w^{\ps} \epsilon^{\pr}{}_{\ps\pt} \omega^{\pt}_{ij}+
2\, {\cal C}_-^{\ps} \epsilon^{\pr}{}_{\ps\pt} \omega^{\pt}_{ij}=0~.
\la{1geom2}
\eea
This is interpreted as a condition which relates ${\cal C}_-^{\ps}$
to the $H_{-ij}$ components of the torsion. As a result, it can be solved to express $H_{-ij}$ in terms of other fields
and the geometry of spacetime.

To determine the conditions imposed on the geometry from the gravitino KSE in directions transverse
to the lightcone, observe that a generic metric connection in 4 dimensions has holonomy contained in
$Sp(1)\cdot Sp(1)$. Thus the only condition required is  the identification of $Sp(1)$ part of the
metric spacetime connection with the $Sp(1)$ part of induced connection from the  Quaternionic K\"ahler
manifold of the hyper-multiplets. This also follows from the integrability conditions (\ref{1int1}).

Thus to summarize, the spacetime admits a null Killing vector field $X$ whose rotation in the directions
transverse to the light-cone is anti-self-dual. The geometry is restricted by (\ref{1geom1}). Furthermore,
(\ref{1geom2}) relates the self-dual $H_{-ij}$ component of the torsion to a component of the induced
$Sp(1)$ connection from the Quaternionic K\"ahler manifold of the hypermultiplets. The metric and torsion
of the spacetime can be written as
\bea
ds^2&=&2e^- e^++\delta_{ij} e^i e^j~,
\cr
H&=&e^+\wedge de^-- \big({1\over16}\omega^\pr_{kl} \nabla_-\omega^{\ps kl} \epsilon_{\pr\ps}{}^\pt
+{\cal C}^\pt_-\big )\,\omega_{\pt ij}\,\, e^-\wedge e^i\wedge e^j
\cr
&&~~~~~~~~~~~~~-{1\over 3!} (de^-)_{-\ell}\,\,\epsilon^\ell{}_{ijk} \,\, e^i\wedge e^j\wedge e^k~.
\la{sumn1}
\eea

The remaining conditions that arise from the KSE are restrictions on the matter content of the theory.
Let us begin with the gaugino KSE. To analyze the conditions, one can choose the gauge
\bea
A_+=0~.
\la{1gaug1}
\eea
In such a case, the components of the gauge connections do not depend on the coordinate adapted to the
Killing vector field $X=\partial_u$. The components $F^{\pa}_{-i}$ are not restricted by the KSE. In the directions
transverse to the light-cone, the self-dual part of $F^{\pa}_{ij}$ is given in terms of the moment maps while
the anti-self-dual part is not restricted. So one can write
\bea
F^{\pa}= F^{\pa}_{-i} e^-\wedge e^i+{1\over2} \mu_{\pr} \omega^{\pr}+ (F^{\rm asd})^{\pa}~.
\eea
This is a Lorentzian version of the Hermitian-Einstein condition.

Turning to the tensorini KSE, it is clear that in the gauge (\ref{1gaug1}), the tensorini scalars
are invariant under the isometries of the spacetime, ie they do not depend on the coordinate $u$.
The 3-form field strengths are self-dual in 6 dimensions. This implies that
\bea
H^\uM_{-\a\b}=H^\uM_{-\a}{}^\a=0~,~~~H^\uM_{-+\bar\a}-H^\uM_{\bar\a\b}{}^\b=0~,~~~H^\uM_{+1\bar 1}-H^\uM_{+2\bar2}=0~,~~~H^\uM_{+1\bar2}=0~.
\la{aself1}
\eea
Combining these conditions with those from the tensorini KSE, one finds that
\bea
H^\uM_{+ij}=0~.
\eea
$H^\uM_{-ij}$ is {\it anti-self-dual} in the directions transverse to the light-cone and the remaining components
are determined in terms of $T$. Therefore
\bea
H^\uM={1\over2} H^\uM_{-ij}\, e^-\wedge e^i\wedge e^j+ T^\uM_i e^-\wedge e^+\wedge e^i- {1\over3!}
T^\uM_\ell\,\epsilon^\ell{}_{ijk}\,\,e^i\wedge e^j\wedge e^k ~.
\la{hum}
\eea

There are some further  simplifications provided we use   (\ref{ric}) to express the KSEs in terms of the fundamental fields. In particular,
(\ref{dpp}) implies that ${\cal C}^\pr_+=0$ and so  (\ref{1geom1}) leads to the geometric conditions
\bea
\nabla_+\omega^\pr=0~,~~~\pr=1,2,3~.
\la{npo}
\eea
In addition, , $T^\uM_i=x^\uM_\ur\partial_i v^\ur$. Substituting this in  (\ref{hum})  most of the components of $H^\uM$ are determined
 in terms of the scalars.
Furthermore, the conditions of the hyperini KSE in the gauge (\ref{1gaug1}) imply that the scalars of the multiplet
are invariant under the action of isometries generated by $X$, ie
\bea
D_+\phi^\ui=\partial_u\phi^\ui=0~.
\la{phiu}
\eea
The remaining restrictions give a holomorphicity-like
condition for the imbedding scalars.

\newsection{N=2 non-compact}

There are two cases with $N=2$ supersymmetry distinguished by the isotropy group of
the Killing spinors. If the isotropy group is non-compact $U(1)\cdot SU(2)\ltimes \bH$, the two Killing spinors
are
\bea
\epsilon_1=1+e_{1234}~,~~~\epsilon_2=i(1-e_{1234})= \rho^1 \epsilon_1~.
\eea
Therefore, the additional conditions on the fields which arise from the second Killing spinor
can be expressed as the requirement that the KSE must commute with the Clifford algebra
operation $\rho^1$.

\subsection{Gravitino}

It is clear that the gravitino KSE commutes with $\rho^1$, iff
\bea
{\cal C}^2={\cal C}^3=0~.
\eea
Equivalently, the gravitino KSE implies that the holonomy of the supercovariant connection is
included in $U(1)\cdot Sp(1)\ltimes \bH$, $\mathrm{hol}({\cal D})\subseteq U(1)\cdot Sp(1)\ltimes \bH$.
The restrictions that this imposes on the geometry will be investigated later.

\subsection{Gaugini}
The gaugini KSE commutes with $\rho^1$, iff
\bea
\mu_2=\mu_3=0~.
\la{2mom}
\eea
These restrictions are in addition to the conditions given in (\ref{1g1}).

\subsection{Tensorini}
A direct substitution of the second Killing spinor into the tensorini KSE reveals that there are no additional
conditions to those given in (\ref{1tens1}). As we have mentioned the tensorini KSE commutes with all $\rho$
Clifford algebra operations.

\subsection{Hyperini}

Combining the restrictions imposed by the second Killing spinor  with those presented in (\ref{1hyp1})
for the first Killing spinor, one finds
\bea
V_+^{\ua\uA}=0~,~~~V_\a^{\ua \uo}=0~,~~~ V_{\bar\a}^{\ua \u2}=0~.
\la{2hyp1}
\eea

\subsection{Geometry}

The form spinor bi-linears are given in (\ref{par1}). The only different is that now the full content of gravitino KSE
can be expressed as
\bea
&&\hat\nabla e^-=0~,~~~\hat\nabla (e^-\wedge \omega)=0~,~~~
\cr
&&\hat\nabla (e^-\wedge \omega^2)-2\, {\cal C} e^-\wedge\omega^3=0~,
\hat\nabla (e^-\wedge \omega^3)+2\, {\cal C} e^-\wedge\omega^2=0~,
\la{2par}
\eea
where we have set $\omega=\omega^1$ and ${\cal C}={\cal C}^1$, ie the form $e^-\wedge \omega$ is covariantly constant
with respect to the connection with skew-symmetric torsion only.

It is clear that the spacetime admits a null Killing vector field $X$, the dual of the 1-form $e^-$, and that
(\ref{dexh}) is valid. The metric and torsion 3-form can be written as in (\ref{1metr}) and (\ref{1tor}),
respectively.

To continue, let us investigate the remaining 3 parallel transport equations in (\ref{2par}). As in the previous
$N=1$ case, the parallel transport equations along the $+$ light-cone direction leads to  (\ref{1geom1}) but with ${\cal C}^2={\cal C}^3=0$.
Thus, one has
\bea
\nabla_+\omega^1_{ij}=0~,~~~\nabla_+\omega^2_{ij}-2\, {\cal C}_+\omega^3_{ij}=0~,~~~
\nabla_+\omega^3_{ij}+2\, {\cal C}_+ \omega^2_{ij}=0~.
\la{p2p}
\eea
The first condition is a restriction on the geometry. The second can be solved for ${\cal C}_+$ to give
\bea
{\cal C}_+={1\over8} (\omega^3)^{ij} \nabla_+\omega_{ij}^2~.
\eea
The third equation in (\ref{p2p}) is automatically satisfied.
Using that $H_{-ij}$ is self-dual and
\bea
\hat\nabla_-\omega_{ij}=\nabla_-\omega_{ij}- H_-{}^k{}_{[i} \omega_{j]k}=0,
\eea
one can solve for $H_{-ij}$ to find
\bea
H_{-ij}=-\nabla_-\omega_{ik}\, I^k{}_j~.
\eea
Two remaining conditions along the $-$ light-cone direction can be used to express ${\cal C}_-$ in terms of the geometry and give some
additional restrictions on the geometry of spacetime.
In particular, one has
\bea
&&{\cal C}_-= {1\over8} \nabla_-\omega^2_{ij} \omega^{3ij}~,
\cr
&&\nabla_-\omega^2_{ij}-\nabla_-\omega^1_{k[i} (I^3)^k{}_{j]}-{1\over4} \nabla_-\omega^2_{k\ell} \omega^{3k\ell}
\omega^3_{ij}=0~,
\cr
&&\nabla_-\omega^3_{ij}+\nabla_-\omega^1_{k[i} (I^2)^k{}_{j]}+{1\over4} \nabla_-\omega^2_{k\ell} \omega^{3k\ell}
\omega^2_{ij}=0~.
\la{2geom1}
\eea

The conditions transverse to the light-cone give
\bea
\tilde H=-i_I\tilde d \omega~,
\eea
where $\tilde d$ is the exterior derivative projected in directions transverse to the light-cone.
This together with the anti-self-duality condition for $H$  turn  (\ref{thde}) into a condition on the geometry
of spacetime
\bea
(de^-)_{-\ell}\, \epsilon^\ell{}_{ijk}=(i_I\tilde d \omega)_{ijk}~.
\la{2geom2}
\eea
 The other two parallel transport equations are automatically satisfied provided that the $U(1)$
part of the curvature tensor of the spacetime connection with torsion is identified with the
  curvature of $U(1)$ connection ${\cal C}$. To see this observe that
 the integrability conditions of the gravitino KSE can be written as
 \bea
 &&\hat R_{\mu_1\mu_2,+\nu}=0~,~~~\hat R_{\mu_1\mu_2,ki}\, I^k{}_j-\hat R_{\mu\nu,kj}\, I^k{}_i=0~,~~~
 \cr
 &&-\hat R_{\mu_1\mu_2,ki}\, J^k{}_{j}+\hat R_{\mu_1\mu_2,kj}J^k{}_{i}-2{\cal F}_{\mu_1\mu_2}
 \omega^{3}_{ij}=0~.
 \la{2int}
 \eea
The second condition implies that the holonomy of the $\hat\nabla$ connection in the directions
transverse to the ligh-cone is contained in $U(2)=U(1)\cdot Sp(1)$. The last condition identifies the
$U(1)$ part of the curvature with the curvature of ${\cal C}$.

To {\it summarize}, the gravitino KSE implies that the metric and torsion can be written as
\bea
ds^2&=&2e^- e^++\delta_{ij} e^i e^j~,
\cr
H&=&e^+\wedge de^--\nabla_-\omega_{ik}\, I^k{}_j\,\, e^-\wedge e^i\wedge e^j
-{1\over 3!} (de^-)_{-\ell}\,\,\epsilon^\ell{}_{ijk} \,\, e^i\wedge e^j\wedge e^k~.
\la{sumn2}
\eea
Of course as in the $N=1$ case, the spacetime admits a null Killing vector field $X$
which also determines components of $H$ and the geometric condition (\ref{1geom1}) is satisfied. Furthermore,
one has to impose the geometric conditions (\ref{2geom1}),  (\ref{2geom2}) and
the restrictions implied by (\ref{2int}).

As we have mentioned the tensorini  KSE does not impose any new conditions on the matter field. As a result,
the restrictions are summarized in (\ref{1tens1}) and the fields are expressed as in (\ref{hum}).

The gaugino KSE gives (\ref{2mom}).  So in the gauge $A_+=0$, one has
\bea
F^{\pa}= F^{\pa}_{-i}\, e^-\wedge e^i+{1\over2} \mu\,  \omega+ (F^{\rm asd})^{\pa}~, ~~~\mu^2=\mu^3=0~,
\eea
where $\mu=\mu^1$.

The hypernini KSE imposes a restriction on the $+$ lightcone direction. The rest of the conditions
 are Cauchy-Riemann type of equations on the scalars.

As in the $N=1$ case, expressing the KSEs in terms of the fundamental fields (\ref{ric}), one can improve somewhat on the solutions
to the KSEs. In particular, the hyperini KSE condition $D_+\phi=0$, ({\ref{dpp}),  implies that ${\cal C}_+=0$. Using (\ref{p2p})
gives rise to the geometric conditions
 \bea
 \nabla_+\omega^1_{ij}=\nabla_+\omega^2_{ij}=\nabla_+\omega^3_{ij}=0~.
\la{2par2}
\eea
Writing $X=\partial_u$ and taking the gauge $A_+=0$, one again concludes that $\phi$ are independent from $u$, (\ref{phiu}).

\newsection{N=2 compact}

\subsection{Gravitino}

The 2 Killing spinors with  isotropy group $Sp(1)$, table 1,  can be chosen as
\bea
\epsilon_1=1+e_{1234}~,~~~\epsilon_2=e_{15}+e_{2345}~.
\eea
The full content of the gravitino KSE is
\bea
\mathrm{hol}({\cal D})\subseteq Sp(1)~.
\eea
The implications that this condition has on the spacetime geometry will be investigated later.

\subsection{Gaugini}

Evaluating the gaugini KSE on $e_{15}+e_{2345}$, one finds
\bea
-2 F_{1\bar2}+\mu^2+i\mu^3=0~,~~~-F_{1\bar1}+F_{2\bar2}+i\mu^1=0~,~~~F_{-i}=0~.
\eea
Combining the above conditions with those  in (\ref{1g1}), we get that
\bea
F^{\pa}_{ab}=0~,~~~ F^{\pa}_{ai}=0~,~~~F^{\pa}_{ij}=- \epsilon_{ijk} \mu^{\pa k}~,~~~a=-,+,\tilde {1}~,
\la{2gau2}
\eea
where $\epsilon_{245}=-1$.
Each of the indices $a$ and  $i$ labels  3 real directions, $i=4,\tilde 2,5$, where we have used
$\tilde 1$ and $\tilde 2$ to distinguish the real directions from the complex directions $1$ and $2$
which naturally appear  in the various conditions which arise from the KSEs. In addition, the $\pr=1,2,3$
index of $\mu$ has been replaced with $k=4,2,5$ after an appropriate adjustment of the ranges and identification
of the components of $\mu$.

\subsection{Tensorini}

A direct substitution of $e_{15}+e_{2345}$ in the tensorini KSE gives
\bea
&&T_-^\uM=0~,~~~H^\uM_{-1\bar1}-H^\uM_{-2\bar2}=0~,~~~H^\uM_{-1\bar2}=0~,
\cr
&&T^\uM_{\bar\a} +{1\over2} H_{-+\bar\a}+{1\over2} H_{\bar\a \b}{}^\b=0~.
\eea
Combining these conditions with those derived for $1+e_{1234}$ and using the self-duality
of $H^\uM$, one finds that
\bea
T_\mu^\uM=0~,~~~H^\uM_{\mu\nu\rho}=0~.
\la{thz}
\eea
So the tensorini KSE vanishes identically. As a result all 8 supersymmetries are preserved. In turn using
the expression of $T$ and $H$ in terms of the physical fields (\ref{ric}), one finds that the scalars
 are constant and 3-form field strengths of the tensor multiplet vanish.

\subsection{Hyperini}

Evaluating the hyperini KSE on $e_{15}+e_{2345}$,  one finds that
\bea
V_-^{\ua\uA}=0~,~~~-V_2^{\ua\uo}+V_1^{\ua\u2}=0~,~~~V_{\bar 1}^{\ua\uo}+V_{\bar 2}^{\ua\u2}=0~.
\la{22hyp1}
\eea
Combining these conditions with those in  (\ref{1hyp1}), we get
\bea
V_a^{\ua\uA}=0~,~~~a=-,+,\tilde{1}~.
\la{2hyp2c}
\eea
The remaining conditions
can be derived by substituting (\ref{2hyp2c}) in either (\ref{1hyp1}) or (\ref{22hyp1}).

Expressing the KSE in terms of the physical fields as in (\ref{ric}), one finds that (\ref{2hyp2c}}) implies
\bea
D_a \phi^\ui=0~,~~~a=-,+,\tilde{1}~.
\la{2hyp2ct}
\eea
The hypermultiplet scalars do not depend on 3 spacetime directions.

\subsection{Geometry}

The algebraic independent form bi-linears are
\bea
e^a~,~~~a=-,+,\tilde{1}~; ~~~e^i~,~~~i=4,\tilde{2},5~,
\eea
where $e^a$ and $e^i$ are 1-forms. The conditions implied by the gravitino Killing spinor equation can be rewritten as
\bea
&&\hat\nabla_\mu e^a=0~,~~~
\cr
&& \hat\nabla_\mu e^i+ 2 \epsilon^i{}_{jk} {\cal C}_\mu^j e^k=0~,
\la{22gravbi}
\eea
where as in the gaugini case the indices $\pr, \ps$ and $\pt$  have been replaced with $i,j$ and $k$,  the ranges
 have been  adjusted, and  the components of ${\cal C}$ have been appropriately identified.
It is clear that the spacetime admits a $3+3$ ``split''. In particular, the tangent space, $TM$, of spacetime decomposes as
\bea
TM=I+\xi~,
\eea
where $I$ is a topologically trivial vector bundle spanned by the  vector fields associated to the three 1-forms $e^a$.

The 1-forms $e^a$ and $e^i$ can be used as a spacetime frame
and write the metric as
\bea
ds^2=\eta_{ab} e^a e^b+\delta_{ij} e^i e^j~.
\eea

Let us first focus on the first equation in (\ref{22gravbi}). This implies that the associated
vector fields to $e^a$ are {\it Killing}.
In addition using the anti-self-duality of $H$, all the components of $H$ can be determined in terms of $e^a$ and its
first derivatives. In particular, one has
\bea
de^a=\eta^{ab}i_b H~,
\eea
where $\eta^{ab}=g(e^a, e^b)$,
and so
\bea
H_{a_1a_2a_3}=\eta_{a_1b}de^b_{a_2a_3}~, ~~~ H_{a_1a_2i}=\eta_{a_1b}de^b_{a_2i}~,~~~H_{aij}= \eta_{ab} de^b_{ij}~.
\la{22hde}
\eea
The first two equations relate the components of $H$ to the commutators of two Killing vector fields
projected along the $e^a$ and $e^i$ directions,  respectively, see \cite{het1,het2}. The anti-self-duality condition for $H$ gives
\bea
H_{a_1a_2a_3} \epsilon^{a_1a_2a_3}=H_{ijk} \epsilon^{ijk}~,~~~
\epsilon_b{}^{a_1a_2} H_{a_1a_2i} =-\epsilon_i{}^{jk} H_{bjk}~,
\eea
where $\epsilon_{013}=\epsilon_{245}=1$.
Thus $H$ can be rewritten as
\bea
H=K-\star K~,~~~K={1\over3!} H_{a_1a_2a_3} e^{a_1}\wedge e^{a_2}\wedge e^{a_3}+{1\over2} H_{i a_1a_2} e^i\wedge e^{a_1}\wedge e^{a_3}~,
\eea
subject to the geometric condition
\bea
(de_{a_1})_{{a_2i_1}} \epsilon^{a_1a_2}{}_{a_3}=-\epsilon_{i_1}{}^{i_2i_3}(de_{a_3})_{i_2i_3}~.
\la{2geom1c}
\eea

Returning to the second equation in (\ref{22gravbi}), one finds that it is equivalent to
\bea
\nabla_b e^i_j-{1\over2} H^i{}_{bj}+2 \epsilon^i{}_{kj} {\cal C}_b^k =0~,
\cr
\nabla_j e^i_k-{1\over2} H^i{}_{jk}+ 2 \epsilon^i{}_{sk} {\cal C}^s_j=0~.
\la{22grav2}
\eea
The first condition again express a component of $H$ in terms of the geometry and ${\cal C}$. Substituting
the expression we have for $H$ in (\ref{22hde}), one finds
\bea
\nabla_a e^i_j+2 \epsilon^i{}_{jk}\, {\cal C}_a^j e^k=-{1\over2}\eta_{ab}\,de^b_{kj}\,\delta^{ki}~.
\la{2geom2c}
\eea
The last condition in (\ref{22grav2}) identifies the spin connection $\hat\Omega$ of the spacetime in directions transverse to the Killing
with the induced $Sp(1)$ connection of the scalars. This can also be seen by looking at the integrability conditions of the gravitino KSE.
In particular,  one has
\bea
\hat R_{AB, aC}=0~,~~~\hat R_{AB, j_1j_2}=-2 {\cal F}^k_{AB} \epsilon_{kj_1j_2}~.
\la{curel}
\eea
These two conditions follow from the integrability conditions
of (\ref{22gravbi}) on $e^a$ and $e^i$, respectively.

Moreover, expressing the KSEs in terms of the physical fields and using  the restrictions imposed
by the gaugini and hyperini KSEs, one also finds
\bea
\hat R_{aB, CD}=0~.
\eea
Similarly, one also has that
\bea
{\cal C}^i_a=0~,
\eea
and so (\ref{2geom2c}) turns into a condition on the geometry.
 It is clear that the only non-trivial components of the curvature
with torsion are those along the transverse to the Killing vector directions and all of them are specified
in terms of the curvature of ${\cal C}$.

To summarize, the spacetime admits 3 Killing vector fields and the torsion $H$ is completely determine in terms of these
and their first derivatives. In particular,
 one has
 \bea
 &&ds^2=\eta_{ab} e^a e^b+\delta_{ij} e^i e^j~,
 \cr
 &&H=K-\star K~,~~~K={1\over3!} H_{a_1a_2a_3} e^{a_1}\wedge e^{a_2}\wedge e^{a_3}+{1\over2} H_{i a_1a_2} e^i\wedge e^{a_1}\wedge e^{a_3}~.
 \eea
 In addition, the spacetime geometry is restricted by (\ref{2geom1c}), (\ref{2geom2c}) and the last condition in (\ref{22grav2}) or equivalently
 (\ref{curel}). The conditions imposed by the remaining 3 KSEs are self-explanatory.

\subsubsection{An example}

Under some additional assumptions, the geometry of spacetime can be described in terms of principal bundles. In particular, one can take
 either that $H$ is closed, $dH=0$, or that the algebra of vector fields associated with $e^a$ closes under Lie brackets.
These two assumptions are related. Following the results of \cite{het3},
if $H$ is closed and the commutator of the vector fields does not close under Lie brackets, then the spacetime
admits  at least an additional parallel vector field. In turn,   the holonomy of the supercovariant connection
reduces to subgroup of $U(1)$. Such solutions admit at least 4 parallel spinors and they are investigated later.
So if one insists on solutions with strictly 2 parallel spinors, $dH=0$ implies that the algebra
of the three isometries closes under Lie brackets.
So  suppose that the algebra of the 3 Killing vector fields closes. In analogy with  the results of \cite{het1},
the spacetime can be thought as a principal  bundle with fibre group which has Lie algebra
\bea
\bR^{2,1}~,~~~\mathfrak{sl}(2, \bR)~,
\eea
where we have used the classification of Lorentzian Lie algebras \cite{medina, josec}.
The closure property of  the Lie algebra of the 3 Killing vector fields requires that
\bea
H_{abi}=0~.
\eea
In turn, the anti-self-duality of $H$ requires that
\bea
H_{aij}=0~.
\eea
In \cite{het1} this component of $H$ was identified with the curvature of the principal bundle.
Thus if $H_{aij}=0$,  the spacetime is locally a product
$G\times \Sigma$, where $G$ is  either $\bR^{2,1}$ or $SL(2,\bR)$ and $\Sigma$ is a 3-dimensional Riemannian
manifold. The curvature of $\Sigma$ is related to the curvature of ${\cal C}$ as in (\ref{curel}). Such a
condition is not trivial as it requires the existence of a metric on $\Sigma$ whose curvature
is equal to a prescribed quantity. A related example is the Calabi conjecture. However there are solutions. For example,
$SL(2,\bR)\times S^3$ is a solution  with the radii of the two factors equal, the
scalars constant, and with vanishing gauge connection.

\newsection{N=4 non-compact}

The Killing spinors are the $Sp(1)\ltimes\bH$-invariant spinors of table 1.  These can be rewritten
\bea
1+e_{1234}~,~~~\rho^1 (1+e_{1234})~,~~~\rho^2 (1+e_{1234})~,~~~\rho^3 (1+e_{1234})~.~~~
\la{4ncks}
\eea
Therefore the KSEs commute with the Clifford algebra operations $\rho^{\pr}$. We shall use this together
with the conditions imposed on backgrounds preserving 1 supersymmetry to derive all the conditions implied by the
KSEs in this case.

\subsection{Gravitino}

The gravitino KSE commutes with the $\rho^{\pr}$ operations iff
\bea
{\cal C}=0~.
\eea
As a result the curvature ${\cal F}$ of ${\cal C}$ vanishes. Thus the full content of the gravitino KSE
can be expressed as ${\rm hol}(\hat\nabla)\subseteq Sp(1)\ltimes\bH$. The restrictions  that this condition imposes
on the spacetime geometry will be examined later.

\subsection{Gaugini}
The KSE commute with $\rho^{\pr}$, iff
\bea
\mu_1=\mu_2=\mu_3=0~.
\la{4mom}
\eea
These are in addition to the conditions given in (\ref{1g1}). Thus, we have that
\bea
F^{\pa}= F^{\pa}_{-i}\, e^-\wedge e^i+ (F^{\rm asd})^{\pa}~.
\eea

\subsection{Tensorini}

The tensorini KSE commutes with the Clifford algebra operations $\rho^{\pr}$. Thus
there are no additional
conditions to those given in (\ref{1tens1})

\subsection{Hyperini}

In addition to the conditions (\ref{2hyp1}), one finds
\bea
 V^{\ua\uo}_{\bar\a}=0~,~~~ V_\a^{\ua\u2}=0~.
  \la{4hyp1}
\eea
Thus the only non-vanishing component is
\bea
V_-^{\ua\uA}
\eea

Imposing the conditions of the hyperini KSE on the physical fields using (\ref{ric}), one finds that the only
non-vanishing derivative on the scalars is
\bea
D_-\phi^\ui~.
\eea
Thus the scalars depend only on one light-cone direction.

\subsection{Geometry}

The spinor bi-linears are the same as those of the $N=2$ non-compact case. The important difference here is that
${\cal C}=0$ and so the conditions imposed by gravitino KSE can be rewritten as
\bea
\hat\nabla e^-=0~,~~~\hat\nabla (e^-\wedge \omega^{\pr})=0~.
\la{4gravbi}
\eea
The solution to these conditions is similar to that of the non-compact $N=2$.
So one  writes
\bea
ds^2&=&2e^- e^++\delta_{ij} e^i e^j~,
\cr
H&=&e^+\wedge de^-- {1\over16}\omega^\pr_{kl} \nabla_-\omega^{\ps kl} \epsilon_{\pr\ps}{}^\pt \,\omega_{\pt ij}\,\, e^-\wedge e^i\wedge e^j
\cr
&&~~~~~~~~~~~~~-{1\over 3!} (de^-)_{-\ell}\,\,\epsilon^\ell{}_{ijk} \,\, e^i\wedge e^j\wedge e^k~.
\la{4sumn1}
\eea
We have used the anti-self-duality of $H$ to relate the $\tilde H$ component to $de^-$ as in (\ref{thde}).

It remains to find the geometric conditions on the spacetime. We have already dealt with the first
condition in (\ref{4gravbi}). To solve the last 3 conditions in (\ref{4gravbi}), one has that
\bea
\hat\nabla_+\omega^{\pr}=\nabla_+\omega^{\pr}=0~.
\la{4geom1}
\eea
This  is a condition on the geometry.
Furthermore, one has that
\bea
\nabla_-\omega^{\pr}_{ij}-H_-{}^k{}_{[i} \omega^{\pr}_{j]k}=0~.
\eea
This together with the self-duality of $H_{-ij}$ can be used to express  $H_{-ij}$  in terms of the geometry as in (\ref{4sumn1}). There are no conditions
on the geometry along this light-cone direction.

Next, the conditions along the transverse to light-cone directions give
\bea
\tilde H=-i_{I^{\pr}} \tilde d\omega^{\pr}~,~~~(\mathrm {no ~\pr ~summation})~.
\eea
Although these may appear as three independent conditions actually they are not. One of them implies the other
two. In turn, this condition together with (\ref{thde}) imply
\bea
 de^-_{-j}\, \epsilon^j{}_{i_1i_2i_3}=(i_{I^{\pr}} \tilde d\omega^{\pr})_{i_1i_2i_3}~,~~~(\mathrm{ no ~\pr ~summation})~.
\eea
This is another condition on the geometry. The restrictions on the fields imposed by the other 3 KSEs have already been
explained.

\subsection{N=3 descendant}

Unlike all other cases, the $N=4$ backgrounds with $Sp(1)\ltimes \bH$-invariant parallel spinors exhibit an
independent descendant with 3 supersymmetries. We have already argued that the conditions on the fields implied by
gravitino, gaugini and
tensorini KSEs remain the same as those for  backgrounds with 4 Killing spinors (\ref{4ncks}). Different conditions
appear only in the analysis of hyperini KSE.

The 3 Killing spinors have been given in (\ref{dessp1}). A direct substitution into the hyperini KSE
reveals that
\bea
&& V_+^{\ua\uA}=0~,~~~  V_\a^{\ua\uo}=V_{\bar \a}^{\ua\u2}=0~,~~~V_{\bar 1}^{\ua\uo}- V_2^{\ua\u2}=0~,~~~V_{\bar 2}^{\ua\uo}+ V_1^{\ua\u2}=0~.
\eea
These conditions are different from those we have found in (\ref{2hyp1}) and (\ref{4hyp1}) which arise for the case
of 4 supersymmetries. It is straightforward to express the above conditions in terms of the physical fields using (\ref{ric}). For example,
it is easy to see that the first condition implies (\ref{phiu}).
The analysis for the geometry of the spacetime we have made in the previous section remains unaltered.
Of course the scalars of the hyperini KSE satisfy different conditions from those of backgrounds with 4 supersymmetries.

\newsection {N=4 compact}

The Killing spinors are  the $U(1)$-invariant spinors of table 1. These can be rewritten as
\bea
1+e_{1234}~,~~~e_{15}+e_{2345}~,~~~\rho^1(1+e_{1234})~,~~~\rho^1(e_{15}+e_{2345})~.
\eea
Thus the conditions on the fields that arise from the KSEs are those we have found for the $Sp(1)$-invariant Killing spinors, and those required for the KSEs to commute with the Clifford algebra operation $\rho^1$.

\subsection{Gravitino}

The Clifford algebra operation $\rho^1$ commutes with the gravitino KSE provided that
\bea
{\cal C}^2={\cal C}^3=0~.
\eea
As in previous cases, the full content of the gravitino KSE can be expressed as ${\rm hol}({\cal D}\subseteq U(1)$. The geometry of spacetime will be examined below.

\subsection{Gaugini}

The gaugini KSE commutes with $\rho^1$ iff $\mu^2=\mu^3=0$. Combining this with (\ref{2gau2}), one finds
\bea
F^{\pa}_{2\bar2}+i\mu^{\pa}=0~,
\eea
where after suppressing the gauge index  $\mu=\mu^1$.

\subsection{Tensorini}

The tensorini KSE commutes with all the Clifford algebra $\rho^{\pr}$ operators. Since both $1+e_{1234}$
and $e_{15}+e_{2345}$ are Killng spinors, one concludes that  all 8 supersymmetries are preserved.  Thus
$T^\uM=H^\uM=0$ as in (\ref{thz}). In turn, the tensorini multiplet scalars are constant and the 3-form field strengths
vanish.

\subsection{Hyperini}
To find the conditions that arise from the hypernini KSE,  one has to simultaneously impose (\ref{22hyp1})
 and (\ref{2hyp1}). Thus one has that
 \bea
 V_a^{\ua\uA}=0~,~~~a=-,+,1,\bar 1~,
 \eea
 and
 \bea
 V_2^{\ua\uo}=V_{\bar 2}^{\ua\u2}=0~.
 \eea
 The only non-vanishing components are $ V_{\bar 2}^{\ua\uo}$ and $V_{ 2}^{\ua\u2}$.

 Using (\ref{ric}), the above conditions can be expressed in terms of the physical fields as
 \bea
 D_a \phi^\ui=0~,~~~a=-,+,1,\bar 1~,
 \eea
 and
 \bea
D_{ 2}\phi^\ui E_\ui^{a1}=D_{\bar 2}\phi^\ui E_\ui^{a2}=0~,
\eea
respectively. Clearly, the scalar fields do not depend on 4 spacetime directions. The last two conditions are Cauchy-Riemann
type of equations along the remaining two directions.

\subsection{Geometry}

A basis for algebraically independent bi-linears is spanned by the 1-forms
\bea
e^a~,~~~a=-,+,1,\bar 1~,~~~e^i~,~~~i=2,\bar 2~.
\eea
The gravitino KSE can be rewritten as
\bea
\hat\nabla e^a=0~,~~~\hat\nabla e^i-2\, {\cal C}\, \epsilon^i{}_j e^j=0~,
\la{4grav4}
\eea
where we have set ${\cal C}={\cal C}^1$.

As in previous cases, the first equation again implies that the vector fields $X_a$ associated with the 1-forms
$e^a$ are Killing
and
\bea
i_a H= \eta_{ab} de^b~.
\eea
It is clear that the spacetime admits a $4+2$ split. In particular, the tangent space $TM=I\oplus \xi$, where
now $I$ is a rank 4 trivial vector bundle spanned by the 4 Killing vectors $X_a$.

The second equation in (\ref{4grav4}) is equivalent to requiring that
\bea
&&(\nabla_a e^i)_j-{1\over2} H^i{}_{aj}-2\, {\cal C}_a\, \epsilon^i{}_j =0~,~
\cr
&&(\nabla_j e^i)_k-2{\cal C}_j \epsilon^i{}_k=0~.
\la{4cgeom}
\eea
In turn, the first  condition in (\ref{4cgeom})  gives
\bea
(\nabla_a e^i)_j-2\, {\cal C}_a\, \epsilon^i{}_j=-{1\over2} \eta_{ab} (de^b)_{kj} \delta^{ki}~,~~~
\la{4xxx}
\eea
as some components $H$ are determined in terms of ${\cal C}$, and both the $e^a$ and $e^i$ bi-linears and their first derivatives.
In addition, $H$ is anti-self dual. This in turn implies that
\bea
&&H_{aij}={1\over3!}\epsilon_{ij}\,\epsilon_a{}^{b_1b_2b_3} H_{b_1b_2b_3}~,~~~H_{a_1a_2 i}={1\over2}\epsilon_{a_1a_2}{}^{b_1b_2} \epsilon_i{}^j  H_{b_1b_2 j}~,
\eea
where $\epsilon_{2\bar2}=i$ and $\epsilon_{-+1\bar1}=i$.
As all components of $H$ are determined in terms of $e^a$ and its first derivative,
 this leads to more restrictions on the geometry of spacetime. These can be expressed as
\bea
&&de^a_{ij}={1\over3!}\epsilon_{ij}\, \epsilon^{ab_1b_2}{}_{b_3} de^{b_3}_{b_1b_2}~,~~~de^{a_1}_{a_2 i}={1\over2} \epsilon_{a_2b_2}{}^{a_1b_1} \epsilon_i{}^j  de^{b_2}_{b_1j}~.
\la{sd4}
\eea
Observe that the rhs of the first equation depends on the structure constants of the algebra
of the 4 Killing vector fields.

The last condition in (\ref{4cgeom}) identifies the spacetime connection along the directions transverse
to the Killing with a $U(1)$ component of the  induced $Sp(1)$ quaternionic K\"ahler  connection. This can also be seen by investigating
the integrability conditions of (\ref{4grav4}). In particular, one finds that
\bea
\hat R_{AB, aC}=0~,~~~\hat R_{\mu\nu, j_1j_2}=-2 {\cal F}_{\mu\nu}\, \epsilon_{j_1j_2}~.
\la{4curel}
\eea
The derivation of these conditions is similar to that of the $Sp(1)$ holonomy case.

There are some additional simplifications provided we use (\ref{ric}) to express the above conditions in terms
of the physical fields. In particular using the hypernini and gaugini KSEs, one finds that apart from (\ref{4curel})
\bea
\hat R_{aB, CD}=0~.
\eea
Similarly ${\cal C}_a=0$ and so (\ref{4xxx}) becomes a condition on the geometry of spacetime.

\subsubsection{Fibration}

The KSEs do not imply that the algebra of 4 Killing vector field closes. Nevertheless, a large class of examples
can be constructed by imposing closure of this algebra. As it has been explained in \cite{het3} and further discussed
in the compact $N=2$ case, if $dH=0$ and one insists in the existence of strictly 4 parallel spinors,
then the algebra of 4 Killing vector fields closes. So the closure of the algebra is a natural assumption to make specially in the
absence of gauge fields. In turn, the closure of the algebra implies
\bea
H_{abi}=0~.
\eea
The Lie algebra of the Killing vector fields must be isomorphic \cite{medina, josec} to one of the following
\bea
\bR^{3,1}~,~~~\mathfrak{sl}(2,\bR)\oplus \mathfrak{u}(1)~,~~~\bR\oplus \mathfrak{su}(2)~,~~~\mathfrak{cw}_4~.
\la{4liealg}
\eea
The spacetime can be interpreted as a principal bundle with fibre group, which has Lie algebra one of those
in (\ref{4liealg}), and base space a 2-dimensional manifold $B$. Moreover it admits a principal bundle connection
$\lambda^a=e^a$ with curvature given by $de^a_{ij}$. Unlike the $N=2$ case, if the fibre group is not
abelian,
the fibre twists over $B$ because of the first equation in (\ref{sd4}). In the abelian case, the spacetime is locally a
 product $\bR^{3,1}\times B$.
Finally the Riemann curvature of  $B$ must be identified with the curvature of the $U(1)$ connection ${\cal C}$.

\newsection{Trivial isotropy group}

Backgrounds with parallel spinors which have a trivial isotropy group admit 8 parallel spinors. The spacetime
is a Lorentzian Lie group with anti-self-dual structure constants. These have been classified in a similar context in
\cite{jose}. In particular,
 the spacetime is locally isometric to
 \bea
 \bR^{5,1}~,~~~AdS_3\times S^3~,~~~CW_6~,
 \eea
 where the radii of $AdS_3$ and $S^3$ are equal, and the structure constants of $CW_6$ are given by a constant self-dual
  2-form
 on $\bR^4$. Moreover
\bea
{\cal F}({\cal C})=0~.
\eea
This concludes the conditions which arise from the gravitino KSE.

The gaugino KSE implies that the gauge field strength vanishes and $\mu^\pr=0$. The tensorini implies that
the 3-form field strengths vanish and the scalars are constants. Similar hyperini KSE implies that the scalars
are constant. In turn using (\ref{ric}), the latter gives ${\cal C}=0$.

\subsection{Descendants}

The case of trivial isotropy group has descendants. In particular, the KSEs allow for backgrounds with
1,2,3 and 4 supersymmetries. However none of them is independent from the backgrounds and their descendants
we have examined in previous cases. The proof of this is required to establish the results outlined in section \ref{d8s}.
 Here we shall
not describe all the steps of the proof. Instead, we shall focus on one case. The rest follow in a similar way. In particular,
let us consider the descendants with 3 supersymmetries  for which the Killing spinors are given in (\ref{des8u1}).
To establish that there are no independent descendants, we have to solve the hyperini KSE for the
spinors given in (\ref{des8u1}). The first two spinors give
\bea
&&V_+^{\ua\uA}=0~,~~~-V_1^{\ua\uo}+ V_{\bar2}^{\ua\u2}=0~,~~~ V_2^{\ua\uo}+ V_{\bar1}^{\ua\u2}=0~,
\cr
&&V_-^{\ua\uA}=0~,~~~-V_2^{\ua\uo}+V_1^{\ua\u2}=0~,~~~V_{\bar 1}^{\ua\uo}+V_{\bar 2}^{\ua\u2}=0~,
\la{4xxy}
\eea
which follows from (\ref{1hyp1}) and (\ref{22hyp1}). Evaluating the hyperini KSE on the third spinor in
 (\ref{des8u1}), one finds
 \bea
&& c_1 V^{\ua\uo}_1+c_1 V^{\ua\u2}_{\bar 2}=0~,~~~-c_1 V^{\ua\uo}_2+c_1 V^{\ua\u2}_{\bar 1}=0~,
 \cr
&& i c_2 V^{\ua \uo}_{\bar 1}-c_3 V^{\ua \uo}_{\bar 2}-i c_2 V^{\ua\u2}_{\bar 2}+c_3 V^{\ua\u2}_{\bar 1}=0~,
 \cr
&& ic_2  V^{\ua \uo}_2+ c_3 V^{\ua\uo}_1+i c_2 V^{\ua\u2}_1+ c_3 V^{\ua\u2}_2=0~.
\la{4xyz}
 \eea
It is clear that if $c_1\not=0$, then the $V$'s vanish and so the hyperini KSE preserves all supersymmetry.
On the other hand if $c_1=0$, it has been argued in section \ref{d8s} that one can always set $c_3=0$.
Setting $c_3=0$ in the last two conditions in (\ref{4xyz}), one finds that
\bea
V_{\bar 1}^{\ua\uo}-V_{\bar 2}^{\ua\u2}=0~,~~~V_2^{\ua\uo}+V_1^{\ua\u2}=0~.
\eea
Comparing this with (\ref{4xxy}), we again find that all $V$'s vanish. Thus again the hyperini KSE preserves
all supersymmetry and so there is not a new descendant.

\newsection{Conclusions}

We have solved the KSEs of 6-dimensional supergravity with 8 real supercharges coupled to any number of vector,
tensor and scalar multiplets
in all cases. For this we have used the spinorial geometry technique of \cite{spingeom} and the similarity of the
KSEs of 6-dimensional supergravity with those of heterotic supergravity. The solutions are uniquely characterized by the
isotropy group of the Killing spinors in $Spin(5,1)\cdot Sp(1)$ as given in table 1.
This is apart from one case where there is an independent
descendant with 3 Killing spinors and isotropy group $Sp(1)\ltimes\bH$, table 2.

The geometry of the solutions depends on whether the isotropy group of the Killing spinors is compact or non-compact.
In the non-compact case, the spacetime always admits a parallel null 1-form with respect to the connection with skew-symmetric
torsion given by the 3-form of the gravitational multiplet. There are backgrounds with $1,2,3$ and $4$ supersymmetries.
The conditions imposed on the fields by the KSEs are given in all cases.

On the other hand if the isotropy group of the Killing spinors is compact, the solutions preserve 2, 4 and 8 supersymmetries.
In the case of 2 supersymmetries,  the spacetime admits a $3+3$ split where the first 3 directions are spanned by
3 parallel vector fields with respect to the connection with skew-symmetric
torsion given by the 3-form of the gravitational multiplet. There is also a natural frame on the spacetime given
by six 1-form spinor bi-linears.  Similarly, the spacetime of solutions with 4 supersymmetries admits a $4+2$ split where
the 4 directions are spanned by 4 parallel vector fields with respect to a connection with skew-symmetric torsion.
The spacetime again admits a natural frame.

In the compact case, the geometry can be further understood provided we take the 3-form field strength
of the gravitational multiplet to be closed or assume that the algebra of the vectors fields constructed
from spinor bi-linears closes. In such a case, the spacetime can be thought of as principal bundle. For solutions
preserving 2 supersymmetries, the spacetime is locally a product $G\times B$, where $G=\bR^{3,1}$ or $SL(2,\bR)$, and
$B$ is a 3-dimensional manifold. For solutions preserving 4 supersymmetries, the fibre group has Lie algebra
$\bR^{3,1}$, $\mathfrak{sl}(2,\bR)\oplus \mathfrak{u}(1)$, $\bR\oplus \mathfrak{su}(2)$ or $\mathfrak{cw}_4$. Moreover unless
the fibre group is abelian, the principal bundle is always twisted over a 2-dimensional base space.

The geometry of 6-dimensional supersymmetric backgrounds is much simpler than those of heterotic supergravity.
The most striking simplification occurs in the analysis of the descendants. There is just one independent
descendant in 6 dimensions as compared to many possibilities that appear in the heterotic case \cite{het2, het3}.
It is therefore likely that all half supersymmetric solutions and supersymmetric near horizon geometries
of 6-dimensional supergravity can be classified
as  similar results have been obtained  for the heterotic supergravity in \cite{newh, hh}, see also \cite{dario}.
However  the presence
of scalar and vector multiplets in 6 dimensions makes the investigation more involved. Usually such proofs
require
some delicate additional information about the couplings of these multiplets. Nevertheless, it is likely
that such analysis can be carried out under some mild assumptions.

\vskip 0.5cm
{\bf Acknowledgments:}~We would like to thank Fabio Riccioni
for many helpful
discussions. GP thanks the Gravitational Physics  Max-Planck Institute at Potsdam for hospitality
where part of this work was done.
 MA is supported by the STFC  studentship grant  ST/F00768/1. GP is
 partially supported
by the EPSRC grant EP/F069774/1 and the STFC rolling grant ST/G000/395/1.

\setcounter{section}{0}
\setcounter{subsection}{0}


\end{document}